\documentclass{aastex}        %finished
\usepackage{graphicx,amssymb,natbib}

\shorttitle{70micron}
\shortauthors{Y. Li}

\begin{document}

\title{Spitzer 70~$\micron$ Emission as a SFR Indicator for Sub--Galactic Regions}
\author{Yiming Li, Daniela Calzetti}
\affil{Dept. of Astronomy, University of Massachusetts, Amherst, MA 01003}
\email{yimingl@astro.umass.edu, calzetti@astro.umass.edu}
\author{Robert C. Kennicutt}
\affil{Institute of Astronomy, Cambridge University, Cambridge, U.K.}
\email{robk@ast.cam.ac.uk}
\author{Sungryong Hong}
\affil{Dept. of Astronomy, University of Massachusetts, Amherst, MA 01003}
\email{shong@astro.umass.edu}
\author{Charles W. Engelbracht}
\affil{Steward Observatory, University of Arizona, Arizona}
\email{cengelbracht@as.arizona.edu}
\author{Daniel A. Dale}
\affil{Dept. of Physics and Astronomy, University of Wyoming, Wyoming}
\email{ddale@uwyo.edu}
\author{John Moustakas}
\affil{Center for Astrophysics and Space Sciences, University of California, San Diego}
\email{jmoustakas@ucsd.edu}

\begin{abstract}
We use Spitzer 24~$\micron$, 70~$\micron$ and ground based H$\alpha$ data for a sample of 40 SINGS galaxies to establish a star formation rate (SFR) indicator using 70~$\micron$ emission for sub--galactic ($\sim0.05-2\ \rm{kpc}$) line-emitting regions and to investigate limits in application. A linear correlation between 70~$\micron$ and SFR is found and a star formation indicator SFR(70) is proposed for line-emitting sub-galactic regions as $\rm \Sigma(SFR)\ ({M_{\odot}\cdot yr^{-1}\cdot kpc^{-2}})=9.4\times10^{-44}\ \Sigma(70)\ \rm{(ergs\cdot s^{-1}\cdot kpc^{-2})}$, for regions with $12+\rm{log(O/H)}\gtrsim8.4$ and $\rm \Sigma(SFR)\gtrsim10^{-3}\ (M_{\odot}\cdot yr^{-1}\cdot kpc^{-2})$, with a 1-$\sigma$ dispersion around the calibration of $\sim0.16$~dex. We also discuss the influence of metallicity on the scatter of the data. Comparing with the SFR indicator at 70~$\micron$ for integrated light from galaxies, we find that there is $\sim40\%$ excess 70~$\micron$ emission in galaxies, which can be attributed to stellar populations not involved in the current star formation activity.
\end{abstract}


\section{Introduction}
\label{intro}

Interstellar dust absorbs UV and optical light and reradiates it at infrared wavelengths. In dusty systems, the use of UV and H$\alpha$ emission to trace recent star-formation is subject to large uncertainties even when dust attenuation corrections are used, since these corrections have large scatter produced by the large range of possible dust content, distribution, and geometry relative to stars and gas present in galaxies \citep[e.g.][]{meurer1999, calzetti2001, kong2004, dale2007, johnson2007, cortese2008}. IR emission then becomes a reliable method to trace star formation rates (SFRs) in galaxies, where the UV light produced by recent star formation is attenuated and reprocessed by dust into the infrared \citep{kennicutt1998}. 

As deep galaxy surveys have often access to limited wavelength information, monochromatic SFR indicators offer advantages over indicators using integrated luminosity over extended wavelength range (e.g., FIR luminosity) by providing `easier to use' recipes. Monochromatic SFR indicators based on infrared emission from both whole galaxies and sub--galactic regions have been investigated in detail, particularly at the wavelength of 8~$\micron$ and 24~$\micron$, by many authors \citep{calzetti2005, wu2005, alonso2006, perez2006, relano2007, calzetti2007, zhu2008, kennicutt2009, rieke2009}, thanks to large samples of nearby galaxies observed at these wavelengths with unprecedented resolution and sensitivity by the Spitzer Space Telescope \citep[e.g.][]{kennicutt2003, dale2009}. Monochromatic SFR indicators at the longer Spitzer bands, 70~$\micron$ and 160~$\micron$ have also been investigated for the integrated light of galaxies \citep{calzetti2010}. These longer wavelengths may provide more reliable SFR indicators than either 8~$\micron$ or 24~$\micron$ as they are close to the peak of dust IR emission \citep{rieke1979, draine2007b, lawton2010}. \citet{calzetti2010}  analyze the emission in those two bands as a SFR indicator using a sample of 189 galaxies, showing that reliable SFR indicators could be established above a SFR $\sim0.1-0.3\ \rm{M_{\odot}\ yr^{-1}}$. \citet{boquien2010} also presented SFR calibration of Herschel (Space Telescope) 100~$\micron$ and 160~$\micron$ bands for spatially resolved regions in M33.

The launch and recent commissioning of the Herschel Space Telescope is providing a sensitive and high angular resolution window into the far-infrared/submm wavelength regime, also tracing the dust peak emission in high redshift galaxies. Upcoming ground facilities at mm wavelengths (ALMA and the Large Millimeter Telescope) will also provide increased sensitivity and coverage for deep surveys in the mm wavelength regime. Such surveys will probe near the rest-frame dust emission peak ($\sim$60-150~$\micron$), at high redshift, e.g., the Herschel Space Telescope will probe the rest-frame dust emission peak up to z$\sim$2. Thus, it is of significant interest to analyze the behavior of the Spitzer 70~$\micron$ band as a SFR indicator in spatially resolved sub--galactic regions, since these regions may resemble actively star-forming and starburst galaxies at high redshift. In this paper, we will investigate the 70~$\micron$ as a SFR indicator in star forming regions, with sizes from $\sim$ 0.05 to 2 kpc.

A study of the 70~$\micron$ luminosity of HII regions has been already presented by \citet{lawton2010} for the Magellanic Clouds. Our analysis differs from that of \citet{lawton2010} in that we probe a large range of galaxies, thus matching the range of properties those studies that have derived SFR(70) for whole galaxies. We also use an unbiased reference SFR indicator proposed by \citet{kennicutt2007, calzetti2007, kennicutt2009}, consisting of H$\alpha$ and 24~$\micron$ emissions. As a draw back, our study will generally probe larger physical scales than the \citet{lawton2010} paper, although our scales are still matched to those of large star formation complexes \citep[$\sim$100 pc,][]{elmegreen2006}.

The paper is structured as follows: \S2 explains the data; \S3 discusses the method for deriving photometry on the sub--galactic regions; \S4 compares the 70~$\micron$ against a reference SFR indicator and then the data are compared with a simple stellar population plus dust model; \S5 provides the calibration of 70~$\micron$ as a SFR indicator. Discussions are presented in \S6 and summary in \S7.

Units in this paper are, $\rm{ergs\cdot s^{-1}}$ for luminosity, $\rm{ergs\cdot s^{-1}\cdot kpc^{-2}}$ for luminosity surface density (LSD, with symbol $\Sigma$), $\rm{M_{\odot}\cdot yr^{-1}}$ for SFR and $\rm{M_{\odot}\cdot yr^{-1}\cdot kpc^{-2}}$ for star formation rate surface density ($\Sigma$(SFR) or SFRD), unless otherwise specified. We adopt a Hubble constant, $H_{0}=70\ \rm{km\cdot s^{-1}\cdot Mpc^{-1}}$. 

\section{Data}

\subsection{Sample Selection and Description}
\label{sample}

Our baseline sample is the SINGS \citep[Spitzer Infrared Nearby Galaxies Survey,][]{kennicutt2003} survey, which obtained images in both the mid-IR and far-IR with the Spitzer Space Telescope, plus ancillary images in the optical. For the present work, we are interested in the subset of images obtained at 24~$\micron$, 70~$\micron$ (with Spitzer/MIPS) and in H$\alpha$. \citet{dale2009, calzetti2010} describe the images, including background subtraction for the H$\alpha$ images, which were obtained from both the KPNO 2.1-m and the CTIO 1.5-m telescopes. Among the 75 galaxies of SINGS, we exclude all ellipticals, S0 galaxies and some irregular galaxies, which satisfy at least one of the following conditions: i) only one central source could be selected but the galaxy is identified as hosting an AGN; ii) there are either bright star(s) across the galaxy disk, or other factors causing the quality of the images, when convolved to the resolution of the 70~$\micron$ images, to be degraded by spurious artifacts. Thus we end up with a sample of 40 galaxies; their galaxy types, nuclear types and adopted distances are listed in Table \ref{tab:sample}. We don't apply any further selection criterion on galaxy properties, and our 40 galaxies cover almost the full range in Hubble types of spiral galaxies and a few irregular galaxies, with different intensities of star formation, from starburst to normal star forming galaxies. Some of these galaxies have or may have an AGN in the center, and these central regions will be excluded from our analysis. What we term as `central regions' are usually the central $\sim$1 kpc of the galaxies, because of the angular resolution ($\sim$ 16'') of the Spitzer 70~$\micron$ images. Hence, when discarding AGN-impacted nuclei, we will also be discarding regions outside of the nucleus that may be affected by the AGN. The large range of properties of the regions matches or exceed (especially in SFR surface density) the range of properties of the whole galaxies in our final sample. The H$\alpha$ images are corrected for [NII] contamination using the [NII]/H$\alpha$ ratios listed in \citet{kennicutt2009}.

\subsection{Oxygen Abundance}
\label{ox}

\begin{figure*}
\center{
\includegraphics[scale=0.75]{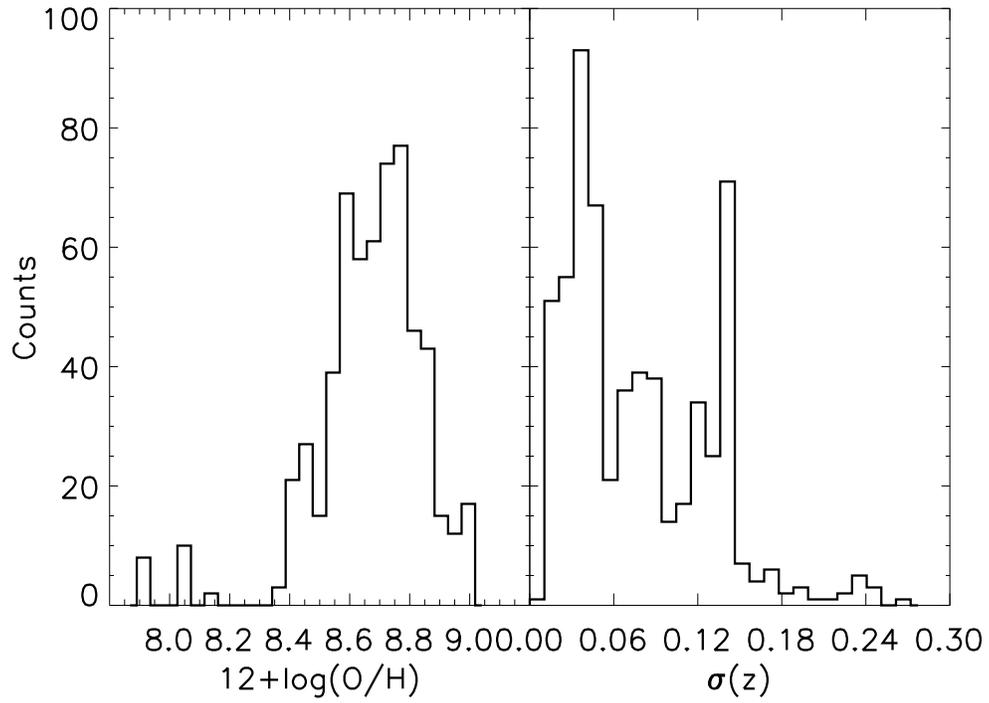}
}
\caption[]{ \centering The distribution of metallicity values for our data on the left panel and metallicity uncertainty on the right panel. The majority are in the intermediate metallicity sub-sample ($8.4<12+\rm{log(O/H)}<8.8$) and have an uncertainty less than 0.15 dex.}
\label{fig:zdis}
\end{figure*}

Oxygen abundances, which we will term `metallicities' in the rest of the paper, for the galaxies in our sample are from \citet{moustakas2010}. They are listed in Table 8 \& 9 in \citet{moustakas2010}; 16 out of 40 galaxies have metallicity gradient measurements. Two calibrations were used by \citet{moustakas2010}, namely KK04 \citep{kk04} and PT05 \citep{pt05}, to derive the metallicity information and we will use the average from these two methods in our analysis. Because of the factor $\sim$5 discrepancy in the metallicity scale resulting from the 2 methods \citep{moustakas2010}, we will avoid, to the extent possible, to refer to absolute metallicity values and mainly use relative values. For those with gradient information, we calculate the metallicity for each aperture taking into account the distance of the center of the aperture to the center of the galaxy, after correcting for the projection effect using inclination information \citep[Table 1 in ][]{moustakas2010}. For the remaining galaxies, we assign the characteristic metallicity of the whole galaxy to each aperture of this galaxy, while for 6 of the small/dwarf galaxies metallicity is derived from the B-band luminosity-metallicity (L-Z) relation and may be susceptible to additional systematics biases \citep{moustakas2010}. Figure \ref{fig:zdis} shows the distributions of the metallicities and of the metallicity uncertainties for the 597 data points in our sample. The uncertainties are derived from the quoted uncertainties in \citet{moustakas2010} in the case of galaxies with directly measured metallicity values or gradients; an uncertainty of 0.2 dex is instead assigned to the metallicity value of galaxies derived from the L-Z relation, reflecting the factor ~5 difference between the two calibration scales.

We divide our sample into three sub-samples based on the adopted metallicities: a low metallicity sample with $12+\rm{log(O/H)}\leqslant8.4$ (`sub-solar', including a total of four galaxies and 41 apertures), an intermediate metallicity sample with $8.4<12+\rm{log(O/H)}<8.8$ ('solar', including a total of 25 galaxies and 425 apertures, where two galaxies' metallicities are from L-Z relation) and a high metallicity sample with $12+\rm{log(O/H)}\geqslant8.8$ (`super-solar', including a total of 11 galaxies and 131 apertures, where four galaxies' metallicities are from L-Z relation). The adopted solar oxygen abundance is 8.69 from \citet{asplund2009}. 

\section{Aperture Photometry}
\label{photometry}

\subsection{Source Selection}
\label{aperture}

\begin{figure*}
\center{
%\scalebox{0.8}[0.8]{\includegraphics{rgb.ps}}
\includegraphics[scale=0.8]{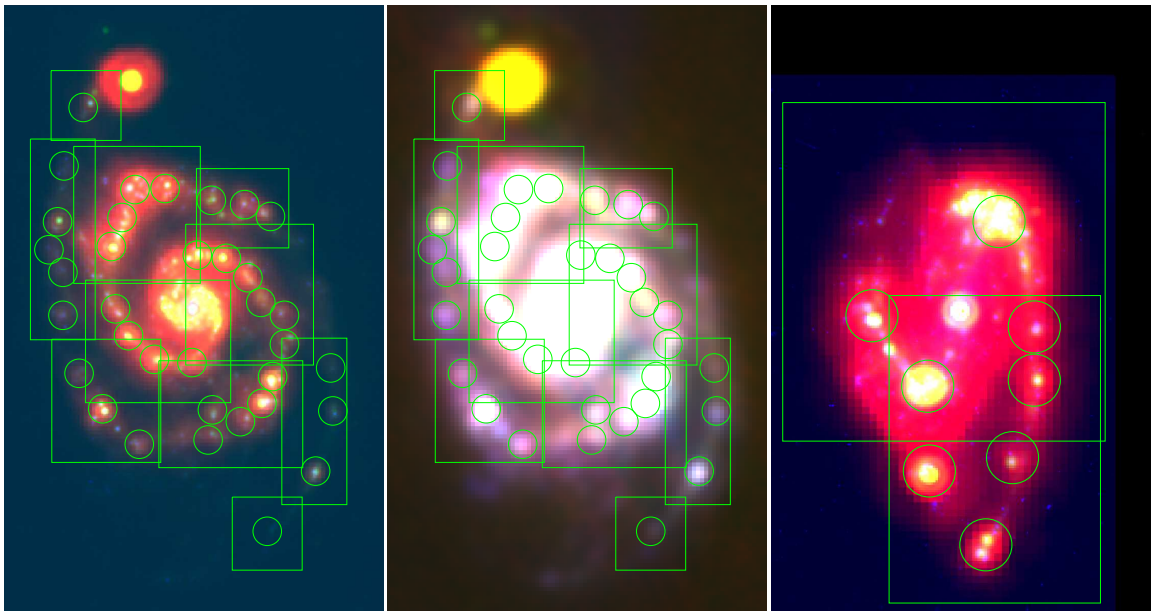}
}
\caption[]{ \centering Aperture and region selection of NGC5194 and NGC3627 on original H$\alpha$ and 24~$\micron$ images (left and right panels), and NGC5194 on convolved H$\alpha$ and 24~$\micron$ images (middle panel). The sizes are $7.05'\times11.25'$ (280$\times$447 pc) for left and middle panels and $3.83'\times6.15'$ (165$\times$265 pc) for right panel. Circles are selected apertures and boxes are regions for local background subtraction. Central regions of both galaxies are not included as both are classified as having an AGN type nucleus. Red: 70~$\micron$; green: 24~$\micron$; blue: H$\alpha$. North is up and east is left. The two galaxies are chosen to show one example of a nearby extended galaxy with multiple resolved features and one example of a farther away, smaller galaxy with a smaller number of identified regions. }
\label{fig:rgb1}
\end{figure*}

For the comparison of multi-wavelength images, all the 24~$\micron$ (6$''$) and H$\alpha$ ($\sim$1$''$-2$''$) images have been convolved to the same resolution as 70~$\micron$ (16$''$) using the convolution kernel and method from \citet{gordon2008} and registered to the same coordinate system and pixel size (4.5 arcsec/pix), after the global background is subtracted, which is determined from the mode of the pixel value distribution of the whole image. The aperture size is chosen to be 16$''$ in radius which corresponds to the FWHM of the 70~$\micron$ PSF (MIPS handbook\footnotemark[1]) and a physical radius $\sim$ 50 pc to 2000 pc depending on the galaxy distance (Table \ref{tab:sample}).
\footnotetext[1]{http://ssc.spitzer.caltech.edu/mips/mipsinstrumenthandbook/MIPS\_Instrument\_Handbook.pdf}
 Sources are selected by manual inspection, at emission peaks of the 70~$\micron$ band; the other two images, at 24~$\micron$ and in H$\alpha$, are then checked for presence of peaks in correspondence of the 70~$\micron$ ones. A candidate within a given aperture is accepted if it appears in all three images (See Fig. \ref{fig:rgb1}). This will bias our sample by excluding very dust obscured objects (i.e. with completely absorbed H$\alpha$) and very transparent objects (with weak IR emission). We don't consider this a major bias in our sample, as \citet{prescott2007} found that in SINGS galaxies only a small fraction ($\sim3\%$) of star forming regions is highly obscured. The exclusion of very transparent objects is also not considered a problem for our analysis, which is centered on the derivation of a SFR calibrator from IR emission, thus requiring presence of dust emission. Although crowding is present within our apertures, and often more than one HII knots are included in them, we usually can identify peaks in each aperture that are brighter than any other in the same aperture. We keep the overlap between apertures to no more than 4\% of the aperture area. Some apertures in crowded environments need to be off-centered because of the overlap criterion and also because of presence of multiple emission peaks within a given aperture. The central regions of these galaxies, which are classified as having or possibly having an AGN by \citet{moustakas2010}, listed in Table \ref{tab:sample}, are not included. With these criteria, we obtain 597 regions out of our sample of 40 galaxies.

\subsection{Local Background Subtraction}
\label{LBG}

Due to crowding, background annuli are difficult to determine for each aperture, without the influence of a neighboring aperture. Thus, we adopt the method of \citet{calzetti2005} to remove the local background from each aperture. Each galaxy in our sample is divided into several local regions usually identified as having a common environment, e. g. within the same spiral arm, after verifying that no sharp decrement of background, caused by either mosaicing problems, other data processing artifacts, or changes in the galaxy's environment, exists within one local background region. Then the local background for each aperture is determined using the mode of the pixel value distribution of the background region. For some more distant galaxies, only a few emission knots are resolved and are quite isolated. Although for these galaxies annuli around photometric apertures would be applicable to remove the background, we still apply our method of local background mode removal, for consistency with other galaxies. A comparison between our method and the standard background annuli method using these isolated regions shows that our method works within 1\% accuracy in removing the background of each aperture. For extremely crowded regions, especially the central regions of most large spiral galaxies, the local background of the region is hard to determine. So higher uncertainty should be expected in those central regions, and extreme caution has been applied in those regions when determining the local background. Although the necessity of performing a local background subtraction is still controversial, it is not much different than a photometric measurement performed with an advanced annuli background to better characterize both the sky background and diffuse emission. We present in Appendix \ref{ap:apc} evidence that local background subtraction is necessary in order to maintain a consistent behavior between low luminosity and high luminosity data, which is essential for performing a reliable analysis.

\subsection{Empirically Established Aperture Correction}
\label{apc}

We then need to define aperture corrections in order to recover the lost flux outside our apertures due to the significant portion of flux contained in the 70~$\micron$ PSF wings. However, the aperture correction value provided by the MIPS handbook is not applicable to our photometry, for several reasons: i) we adopt a local background subtraction derived from the mode of local regions rather than annuli around apertures; ii) there is usually more than one emission peak within each aperture; iii) there is crowding in the aperture. Thus, we have to establish an aperture correction for our case. The use of a PSF at 70~$\micron$ to derive `custom' aperture corrections is also not applicable, since, within each aperture, usually there are several emission peaks as can be seen from the high resolution images (see Fig. \ref{fig:rgb1}), and any aperture correction will need to account for the `extended' nature of our sources. Finally, each aperture suffers from contamination of the PSF wings from neighboring apertures because of crowding. All these make it difficult to establish the aperture correction simply from the theoretical PSF, as it's hard to build a reasonable model of PSF distribution both within and outside the aperture. We thus derive empirical aperture corrections using the original unconvolved high resolution H$\alpha$ images, which we consider to represent the true flux distribution within the aperture we choose; this applies since the aperture size is much larger than the FWHM of the PSF of the original H$\alpha$ images (typically 1$''$-2$''$). We can then compare the original H$\alpha$ images and the convolved H$\alpha$ images to establish median aperture corrections for our aperture photometry (see Appendix \ref{ap:apc}). 

\subsection{Other Corrections and Error Terms}
\label{error}

After the aperture correction (established in Appendix \ref{ap:apc}), the H$\alpha$ photometry is also corrected for [NII] contamination (\S \ref{sample}) and Galactic extinction correction \citep{schlegel1998, odonnell1994}, while the Galactic extinction for 24~$\micron$ and 70~$\micron$ is considered negligible \citep[and references therein]{draine2003}. We do not correct the H$\alpha$ emissions for internal extinction, as we use a combination of H$\alpha$ and 24~$\micron$ in our analysis \citep{kennicutt2007,calzetti2007}. The H$\alpha$ fluxes in this paper are the observed H$\alpha$ fluxes as specified in these articles.

In addition to the uncertainties from global and local backgrounds ($<2\%$), calibration uncertainties (2\% for 24~$\micron$, 5\% for 70~$\micron$ and 10\% for H$\alpha$, SINGS data release guide\footnotemark[2]) and aperture correction uncertainties ($\sim22\%$), we have performed other tests to determine the presence of possible uncertainties introduced by the single convolution kernel temperature used ($<1\%$) and misalignment or misplacing of apertures ($\sim$ 1.5\%). 
\footnotetext[2]{http://irsa.ipac.caltech.edu/data/SPITZER/SINGS/doc/sings\_fifth\_delivery v2.pdf}
The convolution of 24~$\micron$ and H$\alpha$ images to the 70~$\micron$ images resolution uses a convolution kernel, which assumes a blackbody temperature of 50\thinspace K\footnotemark[3]. 
\footnotetext[3]{http://dirty.as.arizona.edu/$\sim$kgordon/mips/conv\_psfs/conv\_psfs.html}
Since \citet{calzetti2000} shows that there are two components of dust, cool ($\sim$ 20\thinspace K) and warm ($\sim$ 50\thinspace K ), we also use kernels of 25\thinspace K and 75\thinspace K to perform the convolution and find that less than 1\% error in measured flux is shown between those two convolution temperature and the 50\thinspace K we use. We also shift our apertures by half pixel (4.5 arcsec/pixel) to see how much the photometry is changed to estimate the error introduced by possible misalignment or misplacing of apertures; this introduces at worst a $\sim$ 1.5\% difference in measured flux.

 These six error terms combined together produce our error estimate, typically $\sim24\%$, for the aperture photometry; we can easily see that the aperture correction uncertainty  at $\sim22\%$ (Appendix \ref{ap:apc}), is the dominant source of uncertainty.

\section{Results and Analysis}
\label{results}

In order to establish the calibration of SFR(70), we first investigate the correlation between 70~$\micron$ emission and SFR and then compare the correlation with a simple model for dust absorption and emission of stellar light.

\subsection{Correlation Between 70~$\micron$ and SFR}
\label{relation}

In order to determine whether the 70~$\micron$ luminosity of sub--galactic line-emitting regions can be used as a SFR indicator and what its limitations may be, we first need a reference unbiased SFR indicator for spatially resolved regions. We intentionally avoid TIR luminosity as the reference SFR indicator because 70~$\micron$ is a major contributor to TIR for most galaxies, since 70~$\micron$ is near the peak of dust emission \citep{rieke1979, draine2007b, lawton2010}. As proposed by \citet{kennicutt2007, calzetti2007, kennicutt2009}, a mixed SFR indicator, involving the combination of an optical and an infrared tracer of SFR,  can provide an unbiased SFR estimate. In this paper, we take the combination, L(H$\alpha$)+0.031L(24), from \citet{calzetti2007}, as our reference SFR indicator. \citet{calzetti2007} derive the coefficient 0.031 from HII regions. The analogous coefficient for the integrated light of whole galaxies, from \citet{kennicutt2009}, is 0.020; the difference between the two is possibly due to the presence of diffuse 24~$\micron$ emission in galaxies \citep{kennicutt2009}. Although our regions span nearly two orders of magnitude in size, the majority of regions are dominated in luminosity by HII regions or clusters of HII regions, also on account of the fact that we remove the local background. The \citet{calzetti2007} calibration is suitable for a continuous star formation history up to 100 Myr \citet{calzetti2010}, and our apertures are estimated to have a median crossing timescale $\sim$ 100 Myr and at most $\sim$ 400 Myr (see next section). The potential non--linearity at $\rm\Sigma(SFR)>0.17\ M_{\odot}\cdot yr^{-1}\cdot kpc^{-2}$ is an important caveat for the application of this calibration \citep{calzetti2007}, but the SFR of our apertures reach that high LSD regime only in 2\% of the sample. Thus the combination from \citet{calzetti2007} is expected to be an appropriate SFR indicator in our case, although we still use in some cases the calibration of \citet{kennicutt2009} for comparison. \citet{calzetti2007} used P$\alpha$ emission as reference SFR to calibrate the unbiased SFR indicator, with 33 galaxies chosen from the SINGS sample, based on the availability of Hubble Space Telescope (HST) P$\alpha$ observations. Their sample and our sample have 19 galaxies in common, and their sample consists mostly late type spiral and irregular galaxies, very similar to the morphology distribution of our sample. Galaxies in the \citet{calzetti2007} sample that are not included in our sample usually show centrally-concentrated (e.g. rings, etc.) star formation, and we had to discard them because of the lower angular resolution of our study coupled with the presence of central AGNs in those galaxies. One galaxy, NGC0024, in our sample, is discarded by these authors due to the quality of P$\alpha$ image \citep{calzetti2007}.

The conversion used to derive SFR is then $\rm SFR (M_{\odot}\cdot yr^{-1}) = 5.45\times10^{-42} L_{intrinsic}(H\alpha) = 5.45\times10^{-42}\ L_{obs}(H\alpha)+0.031L(24) (ergs\cdot s^{-1})$ \citep[with the stellar IMF from][]{kroupa2001,calzetti2010}, and $L(\lambda)=\nu L_{\nu}$ for 24~$\micron$ and 70~$\micron$ following the common definition of monochromatic flux. Adopting a \citet{salpeter1955} IMF in the stellar mass range 0.1-100M$_{\odot}$ would increase the calibration coefficient by a factor 1.51. We use LSDs to eliminate the influence of the galaxy distance uncertainties, similarly to the use of LSDs in establishing the aperture correction.

\begin{figure*}
\center{
\includegraphics[scale=0.9]{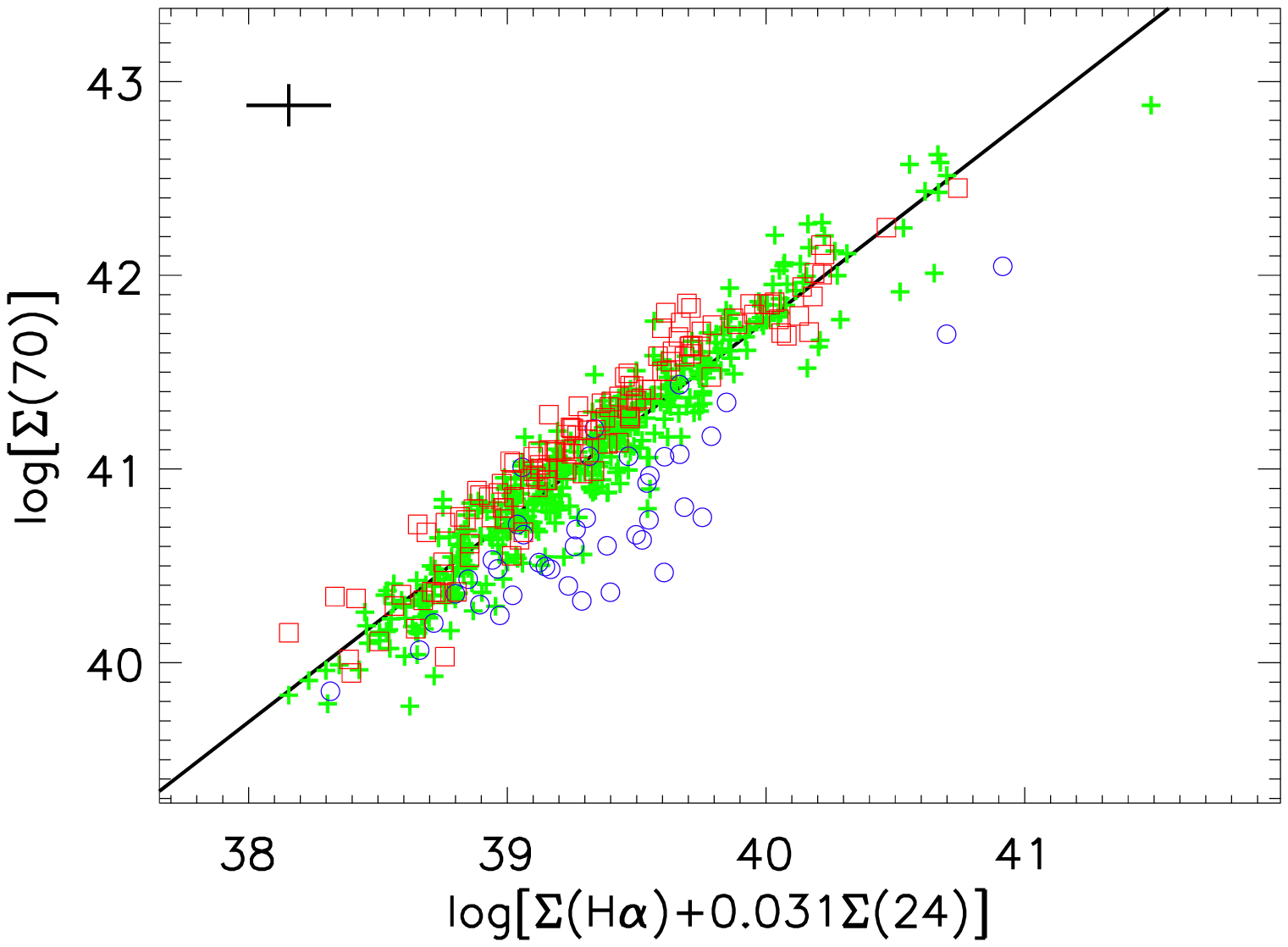}
}
\caption[]{ \centering The LSD of 70~$\micron$ as a function of unbiased SFR indicator (surface density) \citep{calzetti2007} divided into three sub-samples according to the metallicity (Red squares, high metallicity sample with $12+\rm{log(O/H)}\geqslant8.8$; green pluses, intermediate metallicity sample with $8.4<12+\rm{log(O/H)}<8.8$; blue circles, low metallicity sample with $12+\rm{log(O/H)}\leqslant8.4$). The typical error bar is shown as the plus on upper-left. The solid line is the best linear fit through the entire sample. The units are $\rm ergs\cdot s^{-1}\cdot kpc^{-2}$ for LSDs (luminosity surface density).}
\label{fig:dat}
\end{figure*}

The relation between 70~$\micron$ and the reference SFR indicator is shown in Figure \ref{fig:dat}. The linear fit in log-log space gives the correlation between 70~$\micron$ and the reference SFR indicator as
\begin{equation}
\label{equ:fe}
\rm{log}[\Sigma(70)]=(0.342\pm0.504)+(1.036\pm0.013) \rm{log}[\Sigma(H\alpha)+0.031\Sigma(24)]
\end{equation}
The error term in Equation (\ref{equ:fe}) is dominated by the dispersion of the data rather than the uncertainty in each data point. Although there are intrinsic differences between whole galaxies and sub--galactic regions, we also fit for the correlation between 70~$\micron$ and the calibration derived by \citet{kennicutt2009}, for the integrated light of galaxies, as a comparison:
\begin{equation}
\label{equ:k09}
\rm{log}[\Sigma(70)]=(0.184\pm0.526)+(1.042\pm0.013) \rm{log}[\Sigma(H\alpha)+0.020\Sigma(24)]
\end{equation}
This correlation is consistent with the fit in Equation (\ref{equ:fe}) within 1-$\sigma$ uncertainties and both are within $\sim 3-\sigma$ of a linear correlation with a slope of unity between the two quantities.

\begin{figure*}
\center{
\includegraphics[scale=0.9]{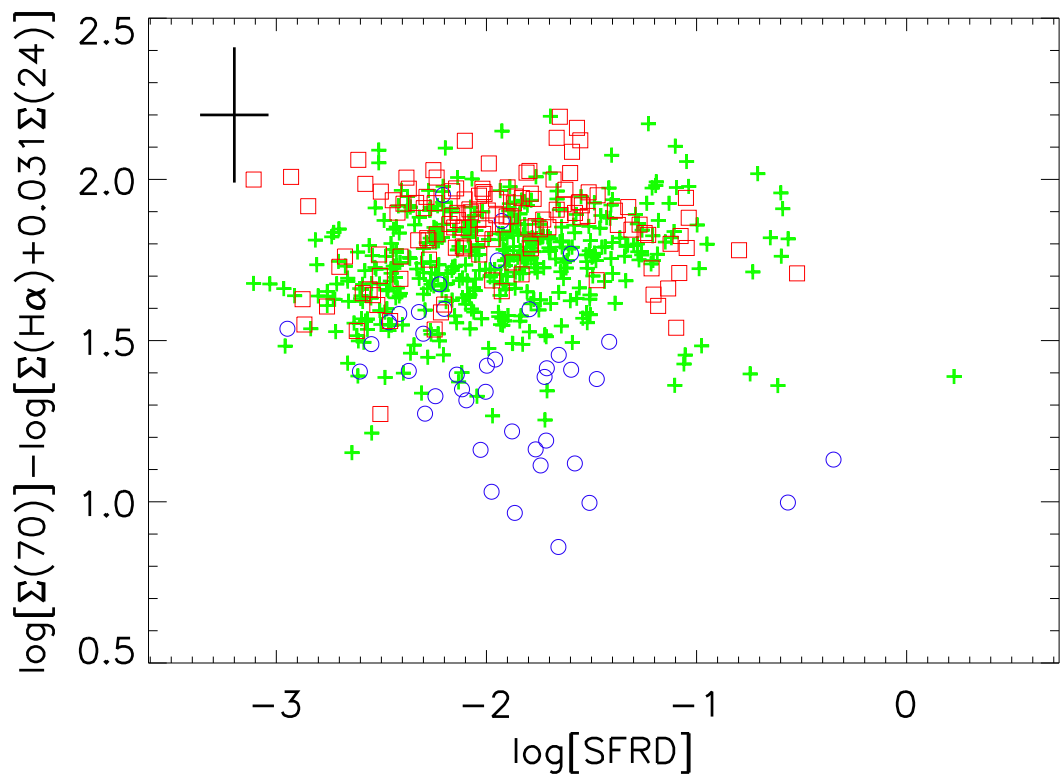}
}
\caption[]{ \centering  The residual between LSD of 70~$\micron$ and unbiased SFR indicator (surface density) as a function of the star formation rate density, SFRD (see caption of Figure \ref{fig:dat} for color-code and symbol-code information). 1-$\sigma$ error is shown as the plus on upper-left for left panel. The units are $\rm ergs\cdot s^{-1}\cdot kpc^{-2}$ for LSDs and $\rm M_{\odot}\cdot yr^{-1}\cdot kpc^{-2}$ for SFRDs respectively.}
\label{fig:extra}
\end{figure*}

\begin{figure*}
\center{
\includegraphics[scale=0.9]{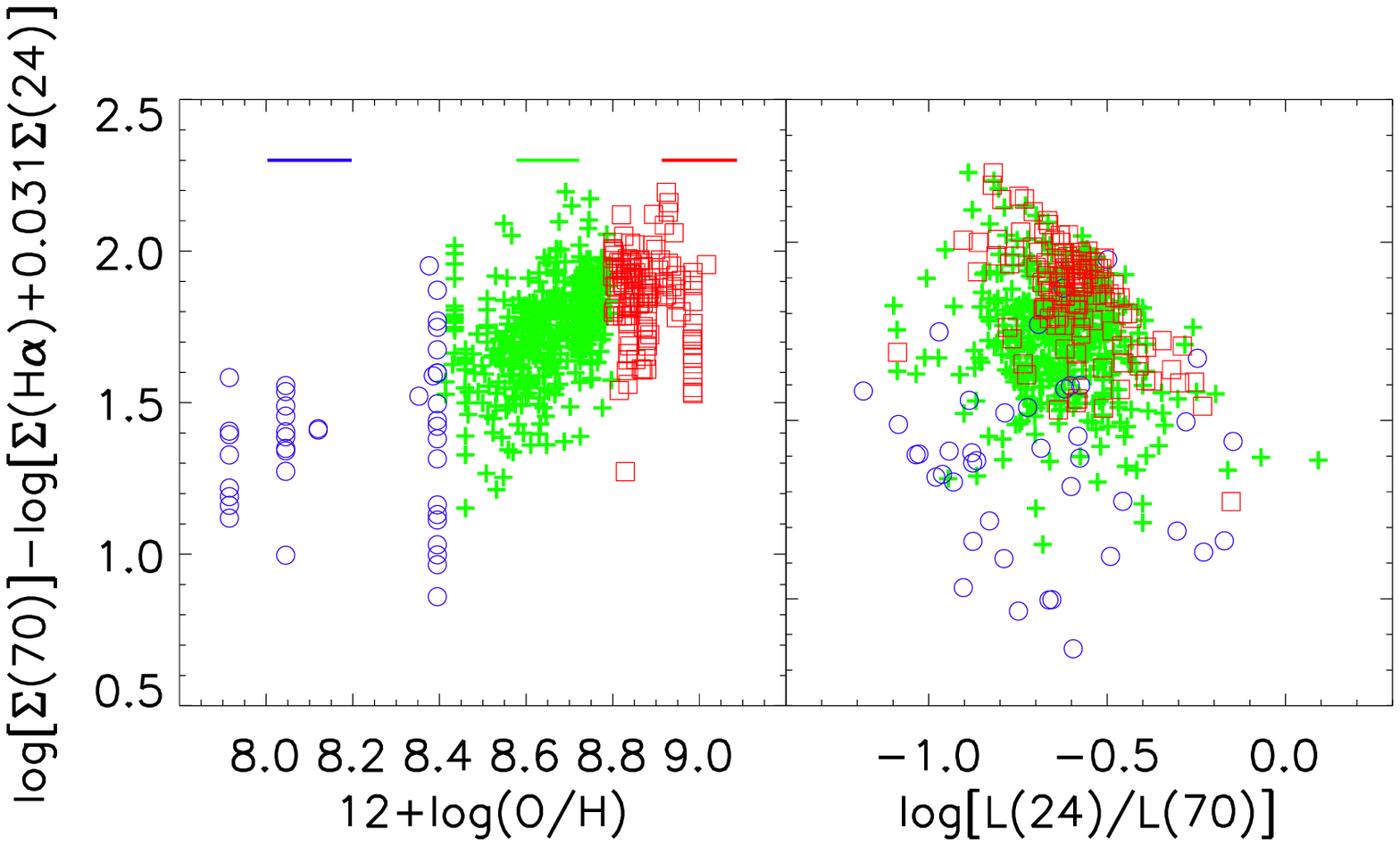}
}
\caption[]{ \centering The residual as a function of metallicity (left hand-side) and L(24)/L(70) ratio (right hand-side, see caption of Figure \ref{fig:dat} for color-code and symbol-code information). The trend is clearly stronger for the residual as a function of metallicity, indicating that variations in the temperature of the warm dust (as traced by L(24)/L(70)) are not the dominant contributor to the scatter around the mean trend. The apparent correlation on the upper right part of the data on the right panel is due to the fact that the plot is actually 1/x versus x once IR emission dominates. Three short lines above the data points in the left panel show the mean metallicity uncertainties of each sample separately. The units are $\rm ergs\cdot s^{-1}\cdot kpc^{-2}$ for LSDs.}
\label{fig:resvt}
\end{figure*}

The slightly steeper-than-unity trend between 70~$\micron$ emission and SFR is a consequence of the increased transparency of the interstellar medium at low SFR \citep{calzetti2010}. At low SFR or low metallicity, the dust has lower opacity which results in a lower IR emission. If the fitting is constrained to high SFR, the slope asymptotically decreases to unity. For example, the slope is $1.004(\pm0.014)$ if fitting is constrained to $\rm log[\Sigma(70)]>40.3$ ($\rm log[\Sigma(H\alpha)+0.031\Sigma(24)]\gtrsim38.9$ or $\rm SFRD\gtrsim0.004\ M_{\odot}\cdot yr^{-1} \cdot kpc^{-2}$), which is consistent with unity. From the 70~$\micron$/SFR vs SFR plot (Fig. \ref{fig:extra}), the ratio of 70~$\micron$ over SFR distributes almost evenly around a constant value; the 1-$\sigma$ dispersion for the whole dataset is $\sim$0.18~dex, and this value decreases to $\sim$0.16~dex if the low metallicity galaxies ($\rm 12+log(O/H)\leqslant8.4$) are removed from the fit. In Figure \ref{fig:resvt}, residuals in the 70~$\micron$/SFR ratio shows the expected correlation with metallicity (i.e. dust content)  and weak or no correlation with L(24)/L(70) ratio (i.e. dust temperature), suggesting that dust content is the major contributor to the systematic scatter around the mean trend although other factors may still produce a small effect. We will compare our data to a simple model in next section to investigate the cause of the scatter in the data. 

\subsection{A Simple Model Analysis}
\label{model}

As can be seen from the left panel in Figure \ref{fig:resvt}, changes in metallicity produce a trend in the 70~$\micron$/SFR ratio, with higher metallicity points displaying higher 70~$\micron$ emission in fixed SFR than lower metallicity ones. We can also see that the low metallicity sample data points, with $12+\rm{log(O/H)}\lesssim8.4$ or $Z\lesssim0.5Z_{\odot}$, show a broader dispersion than the higher metallicity ones in Figure \ref{fig:dat}. At low metallicity, the 70~$\micron$ luminosity could be lower since there is not enough dust to provide the opacity to absorb UV/optical light and reradiate in the IR. However, other factors, such as stellar population age, dust temperature and others could also contribute to the dispersion. Thus, in order to further investigate the nature of the distribution and dispersion of the data, we construct a simple model for dust absorption and emission based on the model of \citet{calzetti2007}. Details of  the model construction are in Appendix \ref{ap:model}, where we use $Z_{\odot}=0.0134$ or $12+\rm{log(O/H)}=8.69$ \citep{asplund2009}.

With the model, we can produce predicted 70~$\micron$ versus H$\alpha$+24~$\micron$ lines for each given metallicity and age. We compare the data with three models of different ages at fixed metallicity and vice versa, and find that the age contribution to the dispersion of data is almost negligible compared to the influence of metallicity. 

\begin{figure*}
\center{
\includegraphics[scale=0.9]{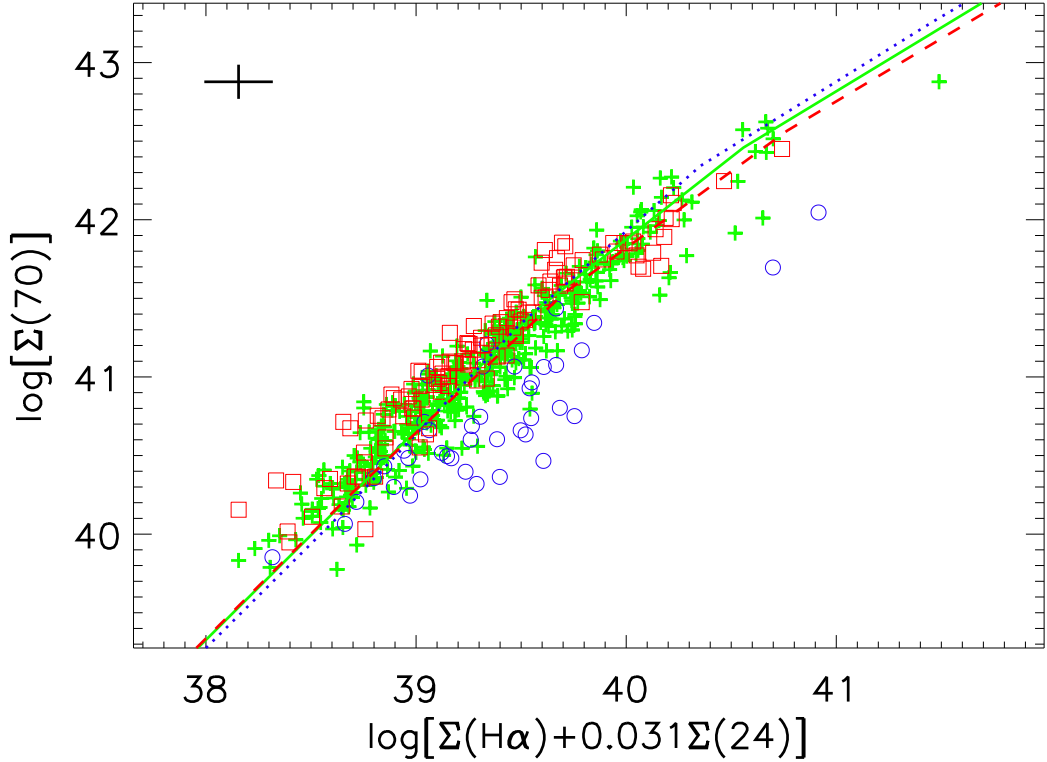}
}
\caption[]{ \centering  Models of 3 different ages, 10 Myr, 100 Myr and 1 Gyr (blue dotted, green solid and red dashed respectively), with solar metallicity, overlaid on data (see caption of Figure \ref{fig:dat} for color-code and symbol-code information).  Typical 1-$\sigma$ error is shown as the plus on upper-left. Different ages of model do not show significant difference from each other. The units are $\rm ergs\cdot s^{-1}\cdot kpc^{-2}$ for LSDs.}
\label{fig:age}
\end{figure*}

In Figure \ref{fig:age}, three models with ages 0.01 (dotted), 0.1 and 1 Gyr (dashed), and about solar ($\sim1.4$ Z$_{\odot}$) metallicity, are overlaid on the data points, the older the redder. If we consider the star formation activity within the region as triggered by a single event, the time required for this perturbation to propagate through out the entire region is comparable to the crossing time ($\sim$ 100 Myr typically for the median 700 pc physical size and from $\sim$ 10 Myr to 400 Myr for all the possible physical sizes, $\sim$ 50 pc to 2 kpc, of our apertures), thus we consider our three age models as bracketing the likely age range of the regions in our sample. The difference between the three models is negligible at the low SFRD end and increases towards high SFRD, but remains small at all SFRs. The small change with age is driven by the fact that, for constant star formation, the amount of UV and ionizing photons (the main contribution to the measured quantities) change negligibly for ages of 10 Myr or longer. Although it is not drawn on the plot, a 10 Gyr model doesn't show a significant difference from a 1 Gyr model but a 2 Myr model does show a significant difference from 10 Myr model at the high SFRD end, as the ionizing population is still growing from 2 Myr to 10 Myr. However, we don't expect the 2 Myr model to be physically applicable in our sample as it is too short compared to the typical crossing time.

\begin{figure*}
\center{
\includegraphics[scale=0.9]{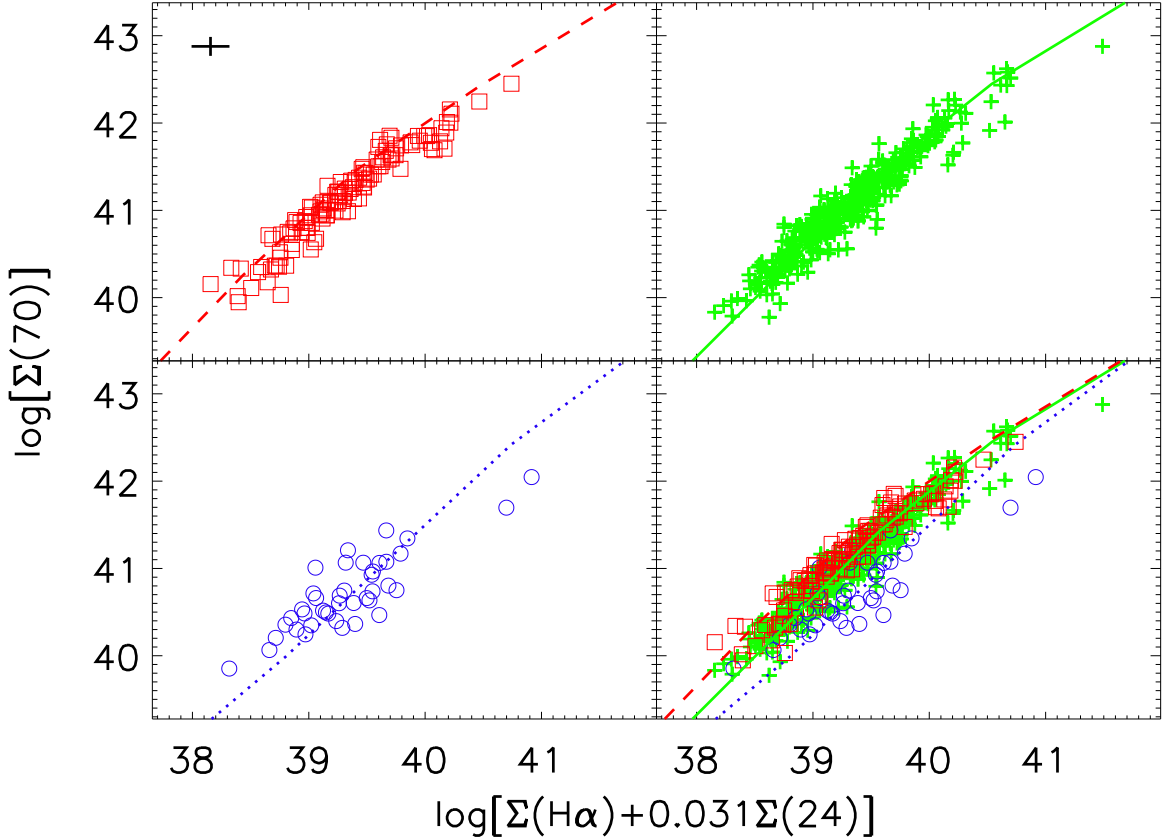}
}
\caption[]{ \centering Three models with different metallicities, 3.7$Z_{\odot}$ (red dashed lines), 1.4$Z_{\odot}$ (green solid line) and 0.3$Z_{\odot}$ (blue dotted line), with an age of 100 Myr, overlaid on data with corresponding metallicity sample (see caption of Figure \ref{fig:dat} for color-code and symbol-code information) separately on each panel and the merged plot on the bottom right panel. The models reproduce the average trend of the data once the different metallicities are taken into account. Combined together, the data show the scatter which we attribute mainly to differences in metallicity. Typical 1-$\sigma$ error is shown as the black plus on upper-left cornel. The units are $\rm ergs\cdot s^{-1}\cdot kpc^{-2}$ for LSDs.}
\label{fig:mod}
\end{figure*}

In Figure \ref{fig:mod}, we keep the age fixed at 100 Myr, of the order of a typical crossing time, and overlay 3 models with different metallicities, 3.7$Z_{\odot}$, 1.4$Z_{\odot}$ and 0.3$Z_{\odot}$, corresponding to the high metallicity, intermediate metallicity and low metallicity samples separately as the values are close to the median of these three sub-samples, on the data. The first three panels show the individual sub-sample overlaid with the corresponding metallicity model line; the models fit the average trend of data pretty well except that there is significant scatter in the low metallicity sample. In the last panel, these three models span a broad range in $\Sigma$(70) at fixed $\Sigma$(SFR), accounting at least in part for the dispersion of the data. This supports our argument that metallicity is the major contributor to the dispersion of the data. The models deviate from a linear relation; this is caused by systematic variations of the average stellar radiation field strength, (U, see Appendix \ref{ap:model}), which increases with increasing luminosity and moves the dust emission peak to shorter wavelengths than 70~$\micron$. The convergence of three model lines at the high luminosity end is caused by the assumption of a scaling between SFR and region opacity, in the sense that more active regions also contain more dust; thus at high luminosity our models simply probe the proportionality between L(70) and L(24), and no longer the scaling with metallicity. Nevertheless, each model is still very close to a linear relationship with unity slope. For decreasing metallicity, the dust infrared emission changes in several ways, which could possibly cause the large scatter in the low metallicity sample. First, a decrease in metallicity will directly reduce the dust opacity, which reduces the extinction and thus the total amount of emission absorbed from stellar UV light. Second, as the metallicity decreases, the effective temperature of dust emission increases \citep{calzetti2000,engelbracht2008}, and thus the trend will start to flatten sooner. However, lacking of sufficient number of low metallicity data could also be an important factor of the data behavior.

\begin{figure*}
\center{
\includegraphics[scale=0.9]{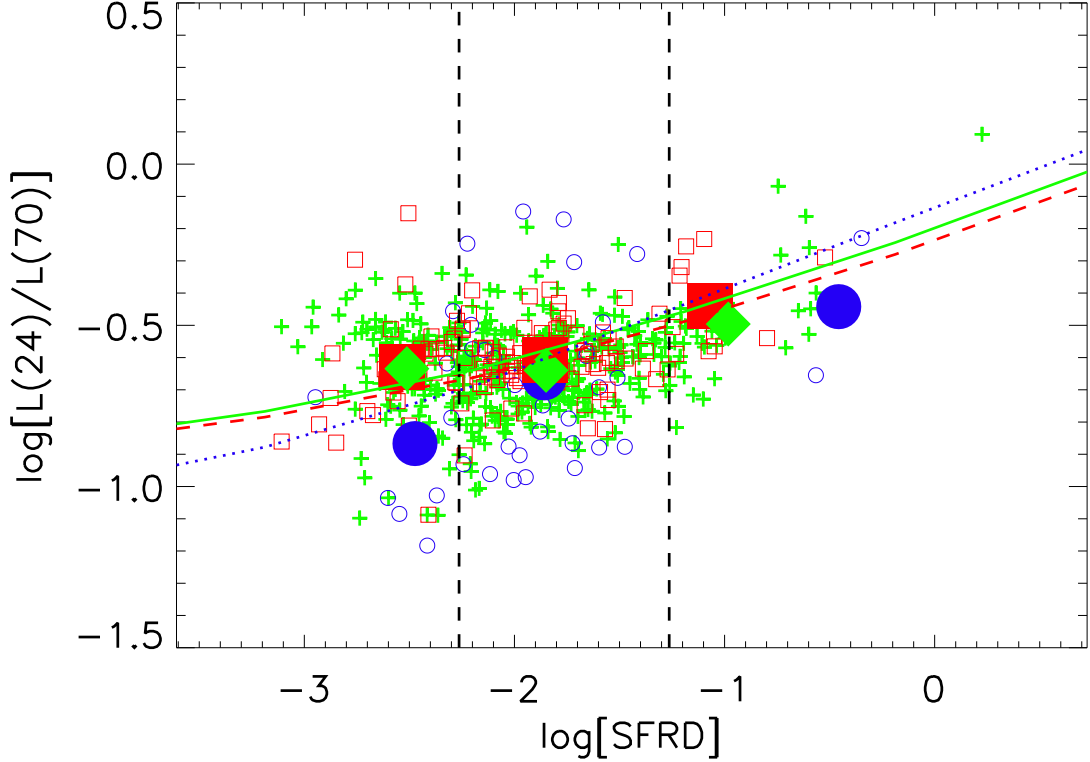}
}
\caption[]{ \centering The luminosity ratio L(24)/L(70) as a function of the star formation rate density, SFRD, for our sample regions (see caption of Figure \ref{fig:dat} for color-code and symbol-code information). Three models with different metallicity (see caption of Figure \ref{fig:mod} for line style information) are overlaid on data. The big filled symbols (blue circles for low metallicity sample, green diamonds for intermediate metallicity sample and red squares for high metallicity sample) represent the binned average (bins separated by vertical dashed lines) for each metallicity sample and the model lines follow reasonably well the average trend of the data. The units are $\rm M_{\odot}\cdot s^{-1}\cdot kpc^{-2}$ for SFRDs and $\rm ergs\cdot s^{-1}\cdot kpc^{-2}$ for LSDs separately.}
\label{fig:rvsfr}
\end{figure*}

To see how the characteristics of the dust emission change as a function of SFRD, we plot the L(24)/L(70) ratio against SFRD, (Fig. \ref{fig:rvsfr}), for both data and models. As can be seen from the plot, the data show a trend that is similar to that of the models, a similarity made even more evident when the comparison is performed with the binned data (filled symbols). The similarity holds at all SFR values, except for the high SFRD bin of the low metallicity sample which suffers from low number statistics; this is in contrast with the results of \citet{calzetti2010} who find that the L(24)/L(70) ratio flattens at high SFRD. We attribute the difference to the fact that the high SFRD data of \citet{calzetti2010} consist mainly of LIRGs, whose dust opacity is sufficiently large that effects of self-shielding of the dust become important and the effective dust temperature no longer increases for increasing SFRD. In our case, regions at high SFRD may still display analogous properties of dust geometry as the low SFRD regions. 

From the comparison between the data and models, and an investigation of the L(24)/L(70) ratio as a function of SFR, we conclude that the observed infrared trends of the HII knots are similar to those of whole galaxies \citep{calzetti2010}, except for the most luminous galaxies. The differences among different HII knots appear to be mainly due to a luminosity scaling, with the more luminous 70~$\micron$ regions being more luminous in all other bands, while the systematic scatter is mostly due to differences in metallicity. This supports a mostly linear dependence between the 70~$\micron$ luminosity and SFR. Because the low metallicity sample shows significantly larger dispersion and deviation from the mean trend, we will exclude the low metallicity sample to derive a SFR(70) calibration.

\section{70~$\micron$ as A SFR Indicator}
\label{SFR}

\begin{figure*}
\center{
\includegraphics[scale=0.9]{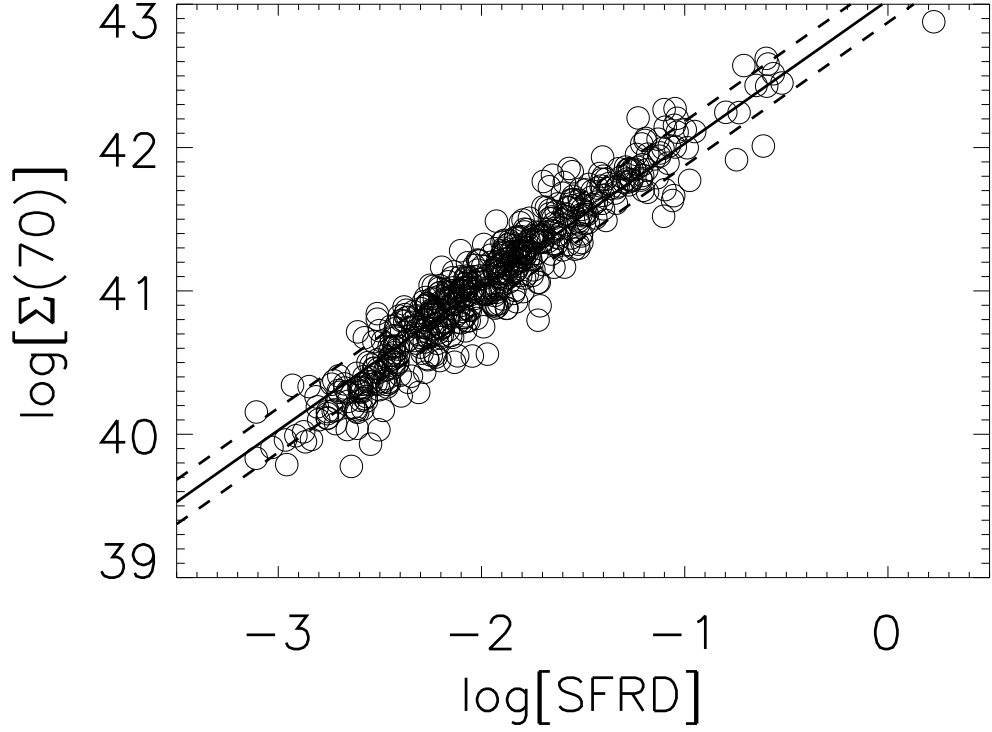}
}
\caption[]{ \centering The data at metallicity larger than $12+\rm{log(O/H)}=8.4$ are fitted with a line with fixed slope of unity in the $\Sigma(70)$--SFRD plane, $\rm log[\Sigma(70)]=log[SFRD]+43.028(\pm0.007)$ (68\% envelop shown as dashed lines). The units are $\rm ergs\cdot s^{-1}\cdot kpc^{-2}$ for LSDs and $\rm M_{\odot}\cdot yr^{-1}\cdot kpc^{-2}$ for SFRDs separately.}
\label{fig:result}
\end{figure*}

We now derive a relation between $\Sigma$(70) and $\Sigma$(SFR) using only data with oxygen abundance greater than 8.4, to reduce the scatter due to the increased transparency of the interstellar medium. The metallicity cut excludes only 4 galaxies and 41 regions, which changes the relation by $\sim7.5\%$. For the remaining 556 regions, spanning almost 5 orders of magnitude in SFRD, we can approximate the trend with a unity slope relation (Fig. \ref{fig:result}),
\begin{equation}
\label{equ:70sfrd}
\frac{\Sigma(SFR)}{\rm{M_{\odot}\cdot yr^{-1}\cdot kpc^{-2}}} = \frac{\Sigma(70)}{1.067(\pm0.017)\times10^{43}\ \rm{erg\cdot s^{-1}\cdot kpc^{-2}}}
\end{equation}
for $2\times10^{40}\lesssim \Sigma_{70}\lesssim5\times10^{42}$.

This translates into a SFR calibration, 
\begin{equation}
\label{equ:70sfr}
\frac{SFR}{\rm{M_{\odot}\cdot yr^{-1}}} = \frac{L(70)}{1.067(\pm0.017)\times10^{43}\ \rm{erg\cdot s^{-1}}}
\end{equation}
for $5\times10^{40}\lesssim L(70)\lesssim5\times10^{43}$. However, the relation between luminosity and LSD has large dispersion due to the uncertainty in the distances. The uncertainty of the calibration coefficient is from the fitting and the dispersion of the data about the mean trend in Equations (\ref{equ:70sfrd}) and (\ref{equ:70sfr}) is $\sim$0.16~dex (dashed lines on Fig. \ref{fig:result}).

Comparing our results with those of \citet{calzetti2010}, we find that our calibration coefficient for the SFR-L(70) relation is 60\% larger than theirs, which means for the same 70~$\micron$ luminosity our calibration will give 60\% higher SFR than the calibration of \citet{calzetti2010}. The difference can be due to the fraction of diffuse emission included in the measurements of whole galaxies, while we probe HII-dominated regions. Since we are only investigating active, star--forming regions, and we remove the local background, we expect a minimal level of contamination from diffuse, non--star--forming stellar populations in our own analysis. 

\section{Discussion}
\label{discussion}

The results and analysis in \S \ref{results} and \S \ref{SFR} give a reliable mean calibration of SFR(70), which can be used under certain luminosity ranges and metallicity limitation (Equation \ref{equ:70sfrd} \& \ref{equ:70sfr}). The origin and impact of the scatter around the mean correlation and the comparison between this calibration with other monochromatic IR SFR calibrations, are interesting and important issues themselves and we further discuss these issues in this section.  

\subsection{Scatters in the Correlation}
\label{scatter}

\begin{figure*}
\center{
\includegraphics[scale=0.9]{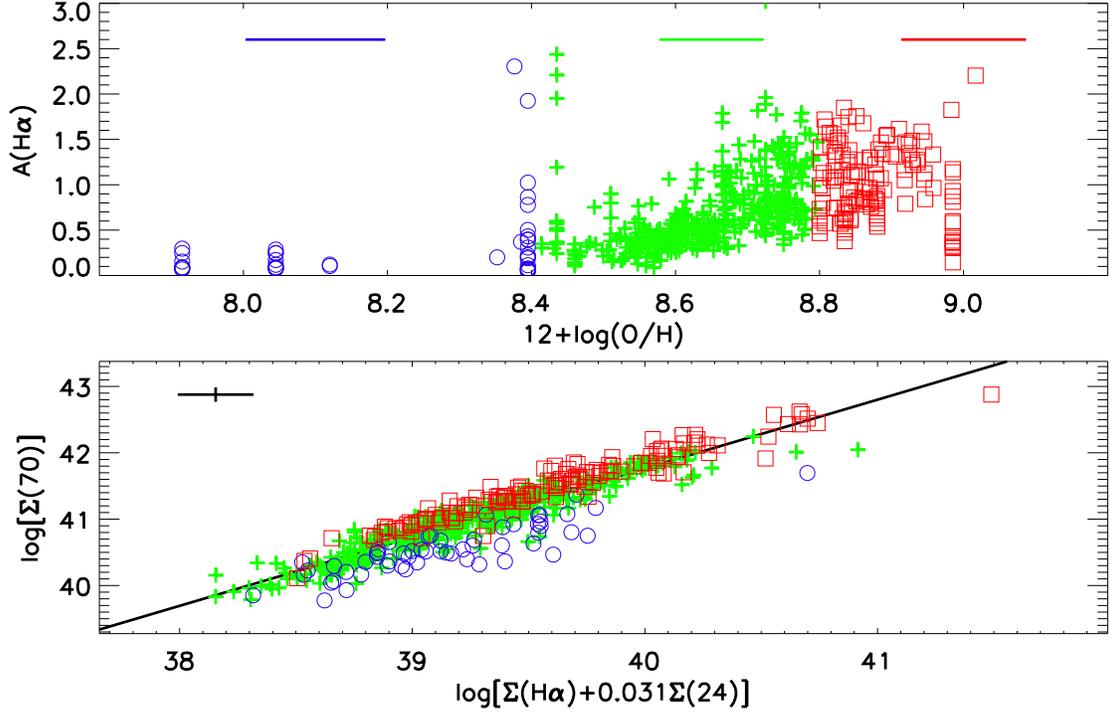}
}
\caption[]{ \centering  On the upper panel, A(H$\alpha$) as a function of metallicity (see caption of Figure \ref{fig:dat} for color-code and symbol-code information). The attenuation of H$\alpha$ is calculated by using the L(H$\alpha$)+0.031L(24) as the intrinsic H$\alpha$ emission \citep{calzetti2007}. The attenuation increases with higher metallicity. The lower panel is the same plot as Figure \ref{fig:dat}, except the data is divided into three sub-samples depending on the A(H$\alpha$) values (red squares, $\rm A(H\alpha)>1$; green pluses, $\rm 0.25<A(H\alpha)<1$; blue circles, $\rm A(H\alpha)<0.25$). However there may be objects with low extinction that have been excluded from the sample, due to our source selection criterion. Three short lines in the upper panel above the data points show the mean metallicity uncertainties of each sample separately. The units are $\rm ergs\cdot s^{-1}\cdot kpc^{-2}$ for LSDs.}
\label{fig:avz}
\end{figure*}

From the analysis of the data and the comparison with the models, we infer that the systematic dispersion in the data around the mean trend is mainly due to variations in metallicity; variations in age and dust temperature can also produce some scatter, but at a much smaller level. Since metallicity and dust attenuation are correlated in first approximation, we should expect a relation between scatter and dust attenuation as well. The upper panel of Figure \ref{fig:avz} shows the attenuation of H$\alpha$ as a function of metallicity: higher metallicity regions do tend indeed to have higher attenuation on average. The attenuation at H$\alpha$ is calculated as the ratio between L$\rm _{obs}(H\alpha)$ and L$\rm _{intrinsic}(H\alpha)$=L$\rm _{obs}(H\alpha)$+0.031L(24), $\rm A(H\alpha)=2.5log[L_{intrinsic}(H\alpha)/L_{obs}(H\alpha)]$\citep{kennicutt2009}. Higher extinction results in more TIR emission, hence more 70~$\micron$ emission; in fact, the lower panel of Figure \ref{fig:avz} shows that the attenuation (red squares, $\rm A(H\alpha)>1$; green pluses, $\rm 0.25<A(H\alpha)<1$; blue circles, $\rm A(H\alpha)<0.25$) traces the scatter similar to that of metallicity. However, there may be objects with low extinction that have been excluded from the sample, due to our source selection criterion.

\begin{figure*}
\center{
\includegraphics[scale=0.9]{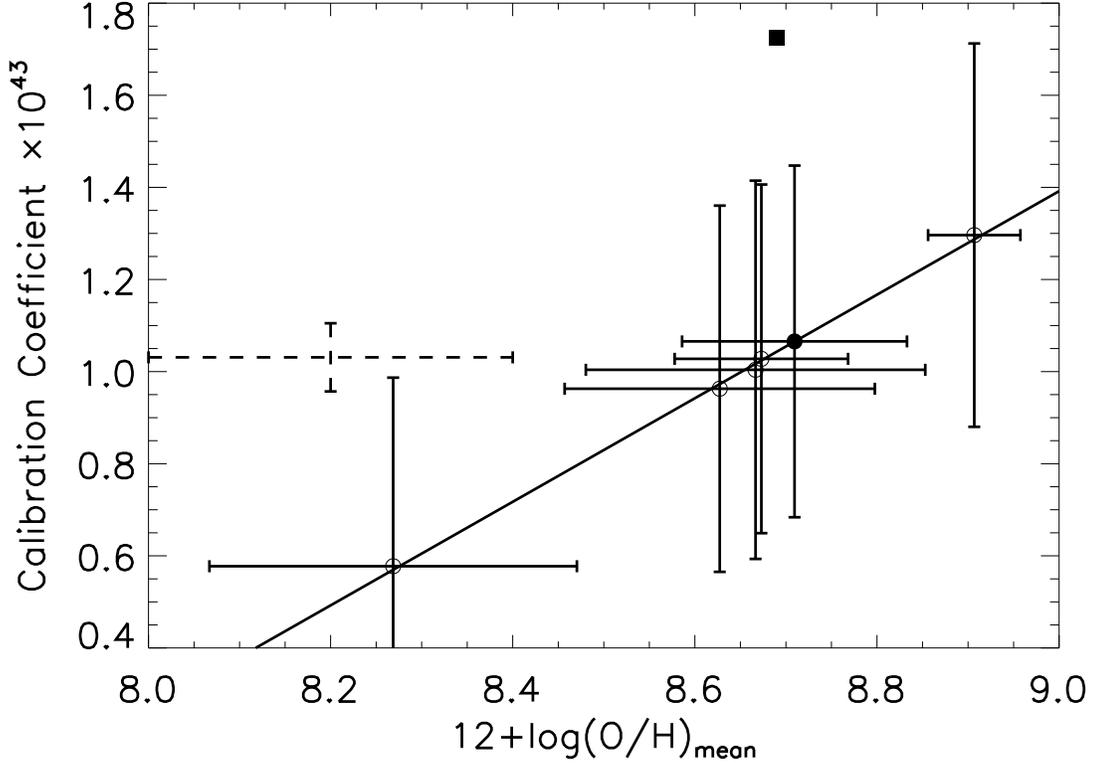}
}
\caption[]{ \centering The calibration coefficient (scaled by a factor of $10^{-43}$) in Equation (\ref{equ:70sfrd}) derived from sub-samples with different (but overlapping) metallicity range as a function of the mean metallicity of the sub-sample. The error bar on mean metallicity shows the standard deviation for each sub-sample metallicity distribution and the error bar on the calibration coefficient shows the 1-$\sigma$ dispersion of each sub-sample. The solid line is a linear fit through our six data points (circles). The filled circle is the mean calibration, adopted in the paper, for the intermediate and high metallicity samples. The filled square is the calibration for integrated 70~$\micron$ from galaxies in \citet{calzetti2010}. The dashed lines indicate the calibration for Magellanic Clouds (taking average of 12+log(O/H)=8.4 for LMC and 12+log(O/H)=8.0 for SMC) from \citet{lawton2010} and the error bar is derived from their dispersion of 70~$\micron$/TIR ratio.These two points are not included in the fit shown in the Figure. }
\label{fig:zdep}
\end{figure*}

As the metallicity systematically introduces a scatter around the mean trend, we should expect that the calibration will change for different metallicity samples and the higher the sample metallicity the larger the calibration coefficient. We attempt to quantify this by dividing our entire sample into six different (but overlapping) sub-samples with different metallicity ranges and deriving the calibration for each sub-sample. The result is shown in Figure \ref{fig:zdep}. The error bar on the calibration coefficient shows the dispersion in data of each sub-sample. From the linear fit on the figure, we could propose a metallicity dependent SFR calibration as
\begin{equation}
\label{equ:zdep}
\frac{SFR}{\rm{M_{\odot}\cdot yr^{-1}}} = \frac{L(70)}{A(Z)\times10^{43}\ \rm{erg\cdot s^{-1}}}
\end{equation}
where $A(Z)=(-8.727\pm9.186)+(1.124\pm1.063)(12+\rm log(O/H))_{mean}$ in terms of mean metallicity of each sub-sample. Our caveat for this calibration is the current unknown nature of the systematic discrepancy in the metallicity values obtained from the KK04 and PT05 \citep{moustakas2010} calibration scales; any change to these scales will change Equation (\ref{equ:zdep}) accordingly.

As a comparison, we plot on Figure \ref{fig:zdep} the calibration constant derived by \citet{lawton2010} for the Magellanic Clouds (with dashed line error bar). The error bar for the Magellanic Clouds calibration is from the dispersion of the 70~$\micron$/TIR ratio in their work. Even though our mean calibration is consistent with \citet{lawton2010}'s result, the two results are slightly inconsistent once the appropriate dependency on metallicity is taken into account. The discrepancy may be due to the reference SFR indicator used by \citet{lawton2010}. These authors use the total infrared emission (TIR), i.e., the dust--absorbed starlight, and the calibration of \citet{kennicutt1998} to derive SFRs for the HII regions of the Magellanic Clouds. However, the SFR(TIR) as derived by \citet{lawton2010} likely underestimates the true SFR in the relatively low--opacity HII regions of the Magellanic Clouds as it only takes account of the obscured SFR and misses the unobscured SFR, thus yielding an overestimated calibration constant for SFR(70). Our test showing a (albeit weak) dependency of the calibration constant on the sub-sample mean metallicity further confirms that metallicity differences do introduce a systematic scatter in the data.

\subsection{Excess 70~$\micron$ Emission in Galaxies}
\label{nsfr70}

From the calibration in \citet{calzetti2010}, a galaxy with a SFR of 1 M$_{\odot}\cdot$yr$^{-1}$ implies a 70~$\micron$ luminosity of $1.725\times10^{43}$ ergs$\cdot$s$^{-1}$. The calibration in this paper shows that resolved HII regions or sub--galactic star forming regions with the same total SFR of 1 M$_{\odot}\cdot$yr$^{-1}$ have a total 70~$\micron$ luminosity of only $1.067\times10^{43}$ ergs$\cdot$s$^{-1}$. The difference in these two calibrations reveals an average of $\sim$40\% excess 70~$\micron$ emission in the galaxies. Both calibrations use the H$\alpha$ emission in their `reference' SFR,  and in \citet{calzetti2010} both the H$\alpha$ emission and the IR emission are measured across the whole galaxy, including any contribution from both the clustered (HII) regions and the diffuse component. Thus, the SFR(70) calibration of \citet{calzetti2010} includes contributions from both components. Conversely, our measurements are local, and explicitly exclude any diffuse contribution, to the extend possible with the angular resolution of MIPS/70; our calibration of SFR(70) thus also excludes any diffuse component from the galaxies. If the heating of the 70~$\micron$ emission in galaxies simply scaled with the H$\alpha$ emission (either clustered or diffuse), then our calibration constant should be the same as that from \citet{calzetti2010} for the same metallicity value. The presence of a significant difference between the two calibration constants (in the sense of `excess' 70~$\micron$ emission in the whole galaxies) suggests that a portion of the 70~$\micron$ emission from whole galaxies is in excess of what can be accounted for from a simple scaling of the H$\alpha$ emission. Hence, we suggest that the excess 70~$\micron$ emission is likely due to stellar populations that are different from those that can ionize hydrogen, i.e., likely to be evolved populations older than about 10 Myr. The only other option is that the photons that ionize the diffuse ionized gas (DIG) can heat the dust more efficiently than in HII regions; we consider this scenario unlikely, as it would require a higher density of such photons than found in HII regions, and this is not observed. Thus, the excess 70~$\micron$ emission should be coming from some `older' or `diffuse' populations. 

Dust heated by older ($>$5--10~Myr), diffuse stellar populations, which are no longer producing ionizing photons and are not related to the most recent star formation activity, are still capable of heating the dust to sufficiently high temperatures that significant emission at 70~$\micron$ can be expected. Small star forming clusters containing only B-- and A--stars (thus, non--ionizing), but not O--stars, could also be partially responsible for the excess 70~$\micron$ emission observed in the integrated light of galaxies. However, we expect the IMF to be fully sampled when averaged over whole galaxies, and thus the effect of smaller star forming regions to be smoothed out.

If we use the SFR calibration of \citet{kennicutt2009} to derive a reference SFR (Equation \ref{equ:k09}), we obtain a calibration coefficient in Equation (\ref{equ:70sfrd}) of $1.285\times10^{43}$, still suggesting that there is $\sim$25\% excess 70~$\micron$ emission in galaxies. The difference in the fraction also suggests a difference in L(24)/L(70) ratio between the galaxies and HII regions, which is discussed in \S \ref{HII2G}. As our regions are only slightly larger than the ones used by \citet{calzetti2007} to establish the reference SFR indicator and also because of the application of local background subtraction, it is reasonable to expect that we should give preference to the calibration of \citet{calzetti2007} for our reference SFR. In summary, the excess light in galaxies at 70~$\micron$ is between 25\% and 40\% of the total, most likely close to $\sim 40\%$.

\begin{figure*}
\center{
\includegraphics[scale=0.9]{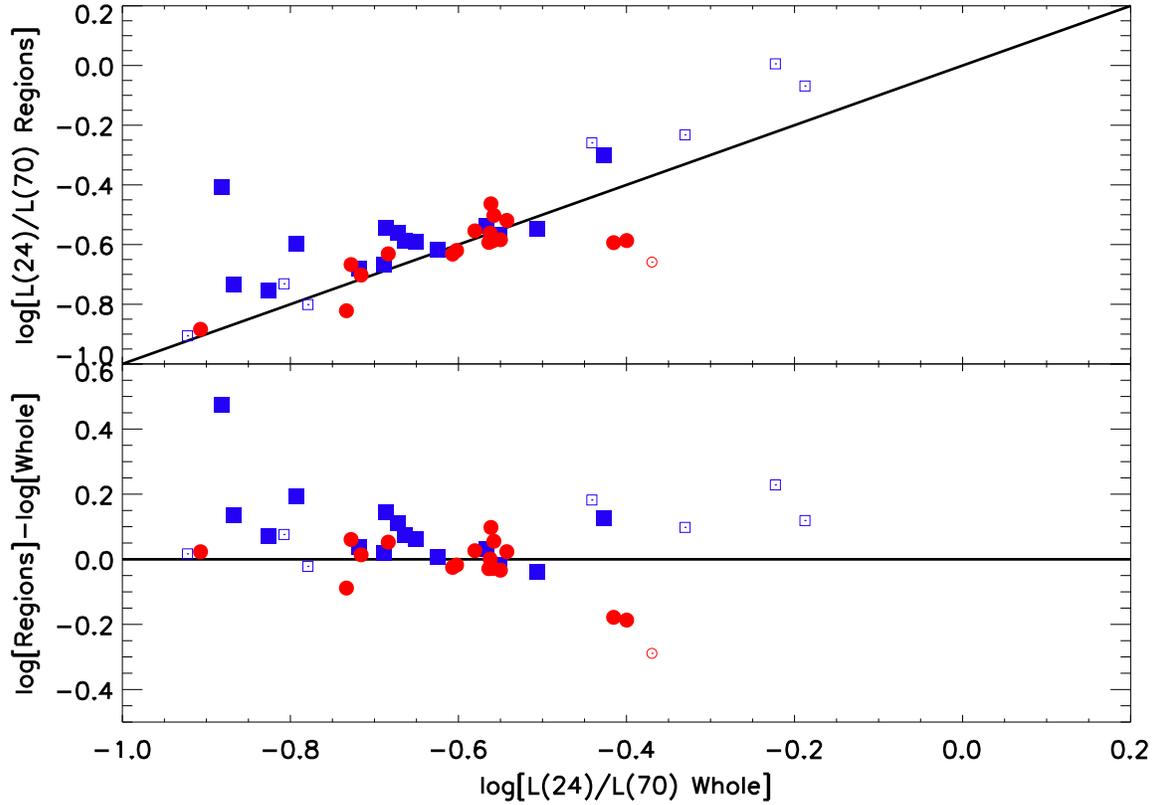}
}
\caption[]{ \centering  The L(24)/L(70) ratio summed up for all the regions in one galaxy as a function of the L(24)/L(70) ratio within the whole galaxy on the upper panel and the residual plot on the lower panel. Red circles are galaxies identified as having an AGN while blue squares are galaxies identified as not having an AGN. Open symbols identify galaxies with less than four regions each. For the AGN-free galaxies, the average dust temperature in HII regions is higher than the galaxy as a whole.}
\label{fig:rvt}
\end{figure*}

The discrepancy between the calibration for galaxies as a whole and for resolved clusters of HII regions is real and significant. In Figure \ref{fig:rvt}, we compare the L(24)/L(70) ratio summed up in the selected regions and that in the whole galaxy for each galaxy. Focusing on the AGN free galaxies (blue squares), the L(24)/L(70) ratio is systematically higher in the line-emitting sub--galactic sources, i.e. the active star-forming regions, than for the integrated light from the whole galaxy, which shows that the dust temperature in star forming regions is higher than that in the whole galaxy  on average. A change in dust temperature for whole galaxies can only be driven by the presence of a stellar population heating the dust to a cooler temperature. We identify this stellar population as `older' and `diffuse'. For the AGN contaminated galaxies (red circles) located below the one-to-one line in Figure \ref{fig:rvt}, the dust could be heated by the central AGN to a higher temperature and also dominate the total IR luminosity.

From all the analysis above, we conclude that there is at least $\sim$25\%, or more likely $\sim$40\%, on average, excess integrated 70~$\micron$ emission from galaxies. This excess comes from dust heated by ÕolderÕ and ÕdiffuseÕ stellar populations, which we identify as stellar populations not related to current star formation activity.

\subsection{Dust Temperature in HII Regions and Galaxies}
\label{HII2G}

From the dust temperature versus SFR comparison (Fig. \ref{fig:rvsfr}), our data follow the models across the full dynamical range, while in \citet{calzetti2010}, a flattening on L(24)/L(70) occurs at the high SFR end. Their high SFR end mostly consists of LIRGs, whose dust distribution becomes optically thick even at IR bands \citep{rieke2009} and the observed dust temperature starts to flatten. The difference indicates that the HII regions in our sample never become optically thick in the IR, even at the highest SFRDs.

\subsection{Comparison with SFR(24)}
\label{24micron}

The 70~$\micron$ emission comprises a large portion of the TIR emission both in galaxies and HII regions/star-forming regions. So a legitimate question is whether the 70~$\micron$ emission is a better SFR indicator than other wavebands. \citet{lawton2010} shows that, in the Magellanic Clouds, the 70~$\micron$ emission is better than 8, 24, or 160~$\micron$ as a SFR indicator, based on the dispersion of the data about the mean relations. Furthermore, \citet{calzetti2007} showed that the relation between the 24~$\micron$ luminosity and SFR is non--linear for HII regions/complexes. In the present work, we find that the 70~$\micron$ emission is linearly correlated with the SFR in HII regions/complexes, and shows almost a factor 2 lower dispersion about the mean trend than the 24~$\micron$ emission \citep[0.16~dex versus 0.3~dex for the 24~$\micron$ emission, ][]{calzetti2007}. Our results thus would seem to support the \citet{lawton2010}'s result that the 70~$\micron$ emission is more tightly correlated with the SFR than the 24~$\micron$ emission in HII regions/complexes. However, we need to caution the reader that the ÕreferenceÕ SFR used here, a combination of H$\alpha$ and 24~$\micron$ is different from the one used in \citet{calzetti2007}, where the extinction-corrected P$\alpha$ is employed.

For whole galaxies, \citet{kennicutt2009} and \citet{rieke2009} show that the 24~$\micron$ emission has a dispersion of only 0.12-0.16~dex about the mean trend with SFR, while \citet{calzetti2010} show that the 70~$\micron$ emission has a larger dispersion, by about 25\%. Whether this indicates that the 24~$\micron$ emission is a better SFR indicator than the 70~$\micron$ emission for whole galaxies is unclear at this stage. Different calibrations rely on different ÕreferenceÕ SFR indicators, and there is a risk of circularity in many comparisons, both for whole galaxies and for HII regions/complexes. A dedicated, independent analysis using a consistent ÕreferenceÕ SFR indicator should be performed to solve this issue.

\section{Summary}
\label{sum}

A sample of 40 galaxies, with high quality H$\alpha$ images and Spizter 24~$\micron$ and 70~$\micron$ images, has been selected from the SINGS legacy survey and 597 sub--galactic regions, in correspondence of peak 70~$\micron$ emission and avoiding AGN contamination, have been identified and measured at H$\alpha$, 24 and 70~$\micron$. For these sub--galactic line-emitting regions (likely groups of HII regions), we have investigated the correlation between 70~$\micron$ dust emission and SFR over scales of 0.5-2 kpc to determine whether we could establish a SFR indicator using the monochromatic 70~$\micron$ emission. We have also investigated dependences on the metal content of the regions, as determined from nebular line emission, via a model constructed with a simple recipe for the stellar population, dust absorption, and emission. For the reference SFR, we have used the combination of observed 24~$\micron$ and the observed H$\alpha$ flux calibrated in \citet{calzetti2007} and \citet{kennicutt2009}. We obtain a relatively tight correlation between 70~$\micron$ and SFR and we provide both a mean calibration and a metallicity dependent calibration. The tight correlation between 70~$\micron$ and SFR is similar to that found by \citet{lawton2010} for HII regions in the Magellanic Clouds, once the difference in physical scale is taken into account. However, our detailed accounting of both the obscured and unobscured SFR in a large variety of galaxies and over a factor $\sim$10 in metallicity enables us to derive a more accurate calibration constant between the 70~$\micron$ emission and the total SFR than done in \citet{lawton2010}. As higher resolution infrared imaging will be obtained by the Herschel Space Telescope in the coming years, this correlation will be further tested with higher spatial detail. Comparing with the SFR indicator at 70~$\micron$ of \citet{calzetti2010}, which is derived for whole galaxies, we find that there is $\sim40\%$ excess 70~$\micron$ emission in galaxies, which we attribute to dust heated by non--star--forming stellar populations. Variations in metallicity in the high and intermediate metallicity sample introduces a dispersion about the correlation of $\sim0.16$~dex but doesn't affect the trend significantly. At the low metallicity end of our sample, the scatter is larger as expected from the lower opacity in the regions. So for deriving the SFR calibration we exclude low metallicity data points. We obtain a mean calibration constant $1.067\times10^{43}$ with a dispersion around the mean trend of $\sim0.16$~dex.

We avoid the regions possibly hosting or contaminated by AGN, so that the infrared emission in our sample is free from AGN contamination. Thus, our SFR relation will not be applicable to sources dominated by AGNs. Our SFR relation is established under the assumption of a universal stellar IMF. Adopting a different IMF will produce a different (scaled) calibration coefficients in Equations (\ref{equ:70sfrd}) \& (\ref{equ:70sfr}).

\acknowledgments

This work is based in part on observations made with the Spitzer Space Telescope, which is operated by the Jet Propulsion Laboratory, California Institute of Technology under a contract with NASA. This work has been partially supported by the JPL, Caltech, Contract Number 1316765.

Yiming Li acknowledges fruitful discussions with and helpful suggestions from Mederic Boquien and Guilin Liu. This work has made use of the NASA/IPAC Extragalactic Database (NED), which is operated by the Jet Propulsion Laboratory, California Institute of Technology, under contract with the National Aeronautics and Space Administration. 

The authors thank an anonymous referee for valuable comments that have helped improve this paper.

\appendix
\section{Empirical Aperture Correction}
\label{ap:apc}

As stated in \S \ref{apc}, we use photometry on unconvolved high resolution H$\alpha$ images to establish our empirical mean aperture corrections for the photometry in 16$''$ apertures on the 24~$\micron$ and 70~$\micron$ images. 

\begin{figure*}
\center{
\includegraphics[scale=0.9]{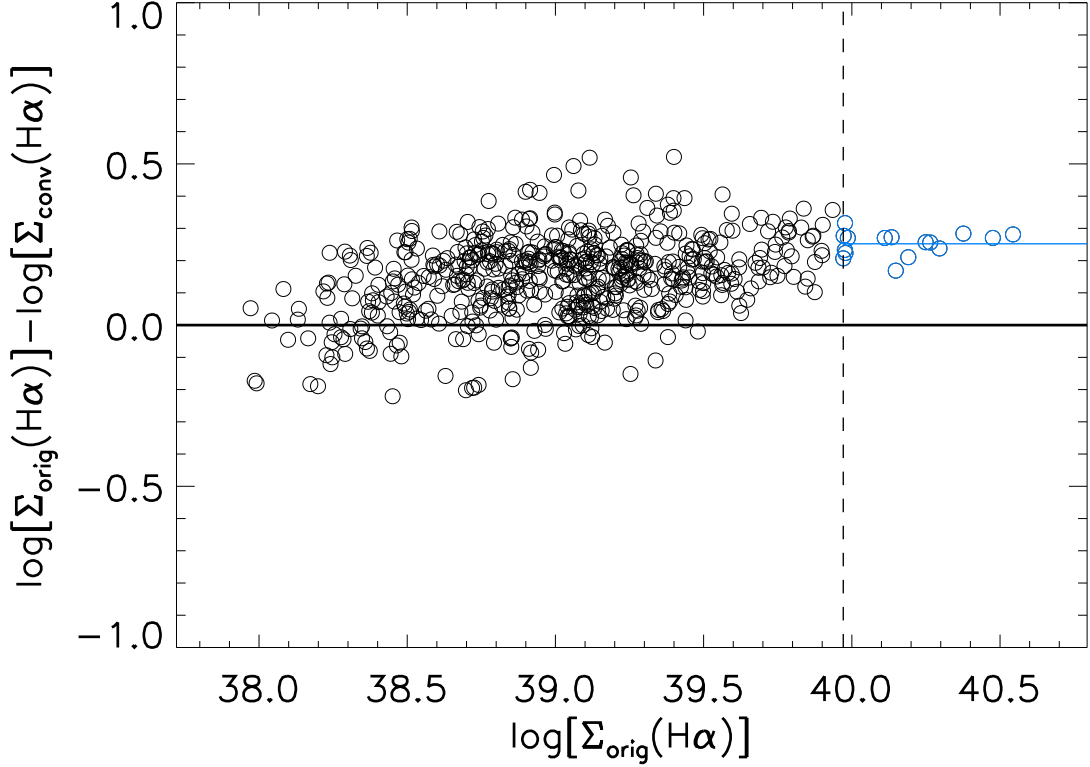}
}
\caption[]{ \centering The residual between original and convolved (to the 70~$\micron$ PSF) photometry as a function of original H$\alpha$ photometry, with local background subtraction. The convolved data are not corrected for any loss outside the aperture which has 16$''$ radius. We use the high end 17 data points ($\sim3\%$,blue points right to the dashed line) of the sample, i.e. the least contaminated apertures, to establish an initial constant aperture correction. The solid blue line indicates the mean of these 17 data points, which is also the adopted constant aperture correction value. The subscripts orig and conv here and in the paper are used to indicate the photometry from the original high resolution images and those from the convolved images respectively. The units are $\rm ergs\cdot s^{-1}\cdot kpc^{-2}$ for LSDs.} 
\label{fig:apc1}
\end{figure*}

Figure \ref{fig:apc1} shows how the difference between the original unconvolved photometry and convolved photometry is distributed as a function of the original photometry of H$\alpha$ as a reference (with local background subtraction). If we had one perfect point source in each of our apertures, the expected correction to the flux would be 0.3~dex (from the MIPS handbook, using background annuli from 18$''$ to 39$''$). Most of the data distribute a little below the horizontal line of $\sim$ 0.3~dex; this is expected since a method for local background removal (\S \ref{LBG}) makes the aperture correction smaller as it receives a smaller contribution from the PSF wings than the background annuli method used in the MIPS handbook. For decreasing luminosity we can also see that the difference tends to be smaller or even below 0, which means the convolved photometry is getting larger than the original one. This trend is due to the contamination of neighboring apertures, and the fainter the aperture the higher the contamination. For the extremely faint apertures, the trend flattens, as these regions can only be `identified' in relatively uncrowded regions. Since within a given aperture there is more than one emission peak for most cases, most apertures are not perfectly centered on the emission peak. This will produce a dispersion in the aperture correction due to different distribution of emission peaks within a given aperture. Those few data points indicating aperture corrections greater than a factor $\sim2.5$ (Fig. \ref{fig:apc1}) are due to the relatively faint emission within the regions, because either the S/N is low or the aperture is substantially off-centered to avoid overlapping with adjacent bright sources. This is also reflected in the large error bars in the photometry of these data points. In summary, the trend of Figure \ref{fig:apc1} indicates that in addition to the overall constant aperture correction due to the PSF wings loss outside our selected apertures (observed in the high luminosity region of the plot) there is also a surface brightness dependent aperture correction, indicative of the contamination effects of neighboring apertures.

\begin{figure*}
\center{
\includegraphics[scale=0.9]{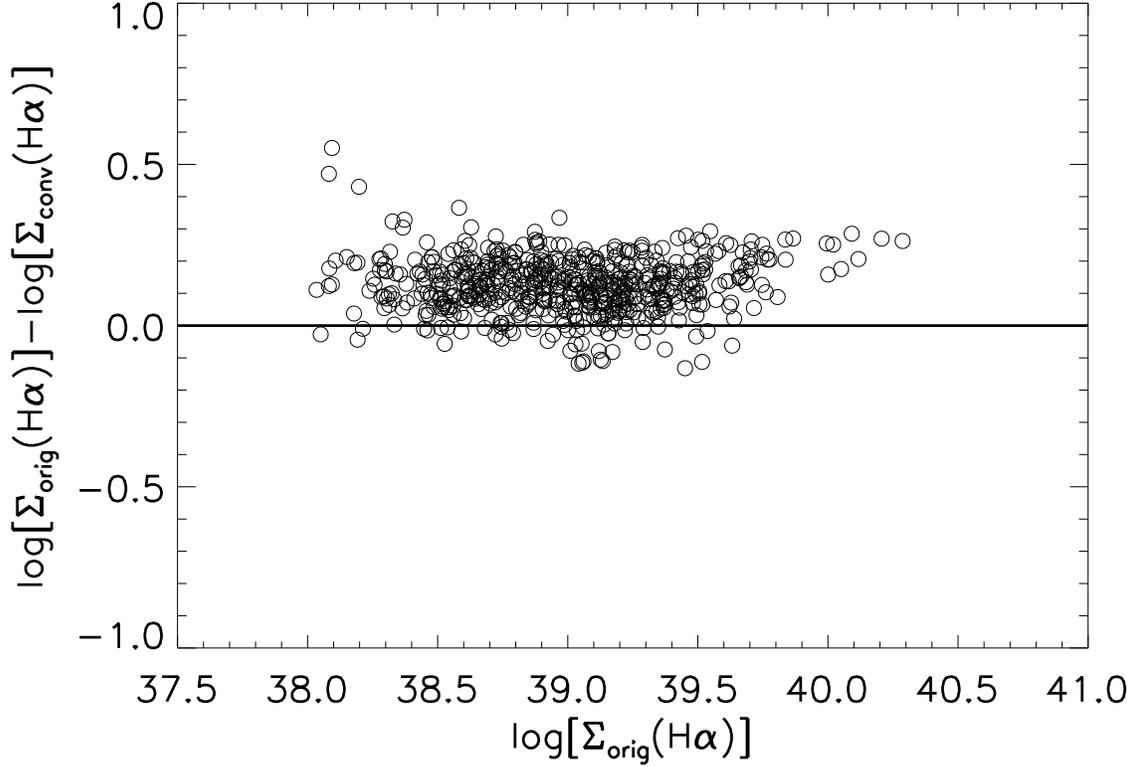}
}
\caption[]{ \centering The residual between original and convolved photometry as a function of original H$\alpha$ photometry, without local background subtraction for both. The low end shows an apparent flaring trend compared to the local background subtracted one, which suggests the need to apply the local background subtraction, especially for low luminosity data points. The units are $\rm ergs\cdot s^{-1}\cdot kpc^{-2}$ for LSDs.} 
\label{fig:nlbg}
\end{figure*}

Figure \ref{fig:nlbg} shows the same plot without local background subtraction for either photometric measurement. The significant difference in this figure relative to Figure \ref{fig:apc1} is the flaring of the data points at the low end of the luminosity distribution, showing a markedly different behavior from the high end. Furthermore, the low-luminosity end is still more than 0.1~dex, on average, below the mean of the high-luminosity end in Figure \ref{fig:nlbg}, again showing evidence for contamination from neighboring sources. We interpret the differences and similarities between Figure \ref{fig:apc1} and \ref{fig:nlbg} as indicating the necessity of subtracting a local background from all apertures and the photometry measurements used hereafter are all with local background subtraction.  

\begin{figure*}
\center{
\includegraphics[scale=0.9]{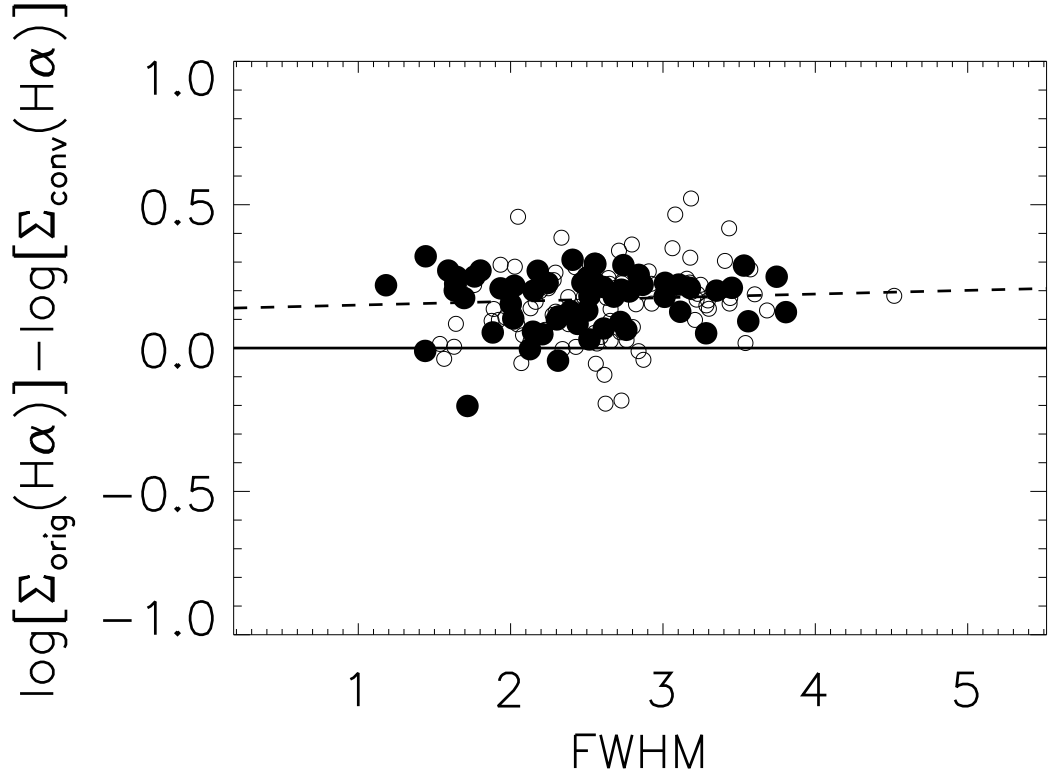}
}
\caption[]{ \centering The residual between original and convolved photometry as a function of the FWHM of the point source within the aperture. Filled circles are those with only one bright point source, contributing at least 50\% of the flux within 2FWHM area, and open circles are those with one brighter point source but also a companion weaker point source, where the FWHM of the brighter source is measured. The dashed line is a linear fit to the filled circles. The aperture correction is not correlated with the compactness of the source, represented by the FWHM, within the apertures. The units are arcsec for FWHM and $\rm ergs\cdot s^{-1}\cdot kpc^{-2}$ for LSDs separately.} 
\label{fig:fwhm}
\end{figure*}

We finally test our aperture corrections for an additional source of bias: the impact of sources compactness. We can technically argue that a more compact source will show a larger aperture correction than a more diffuse source, when convolved to lower resolutions. Thus, our `decrease' in aperture correction with decreasing luminosity may simply indicate that less luminous sources are more diffuse than more luminous sources. To test for this effect we identify within our apertures all those that are dominated by a single source, i.e. the sources represent more than half of the flux in the entire aperture, and measure their FWHM in the original unconvolved images (58 sources). We also identify an additional 96 apertures where two sources dominated the flux but one is significantly brighter than the other. Figure \ref{fig:fwhm} shows the first group of 58 sources as filled circles, and the second group as empty circles. We can see that both groups do not show a dependence of the aperture correction on FWHM or the compactness of the sources within each aperture. A linear fit (dashed line) through the 58 filled circles, the clean sample, shows consistency with a slope equal to zero. We thus conclude that our original interpretation, the decreasing aperture correction with decreasing luminosity indicates increasing contamination from neighboring sources, is  the most likely. We will then use luminosity surface density to quantify aperture corrections for our data.

Based on the discussions above, we model our aperture correction as two parts, a constant aperture correction for recovering the loss of PSF wings and a surface brightness dependent aperture correction for removing the contamination of PSF wings from neighboring apertures.

\subsection{Constant Aperture Correction}
\label{consapc}

Since the most luminous apertures are proportionally less affected by contamination from neighboring apertures, we use the high end of Figure \ref{fig:apc1} as a gauge to establish the constant aperture correction. From Figure \ref{fig:apc1} we can see that the 17 high LSD apertures, $\rm{log}[\Sigma_{orig}(H\alpha)]\gtrsim40$, more or less lie along one horizontal line with some dispersion. So we take the low end, 39.971, of these 17 points, $\sim3\%$ of the total data (blue points on Figure \ref{fig:apc1}), as the cut off LSD and we take the mean aperture correction of these points and get an aperture correction as 1.788. The choice of LSD cut-off is arbitrary, but it won't change the final correction much as long as it stays in the high luminosity end.

\subsection{Surface Brightness Dependent Aperture Correction}
\label{Ldepapc}

\begin{figure*}
\center{
\includegraphics[scale=0.9]{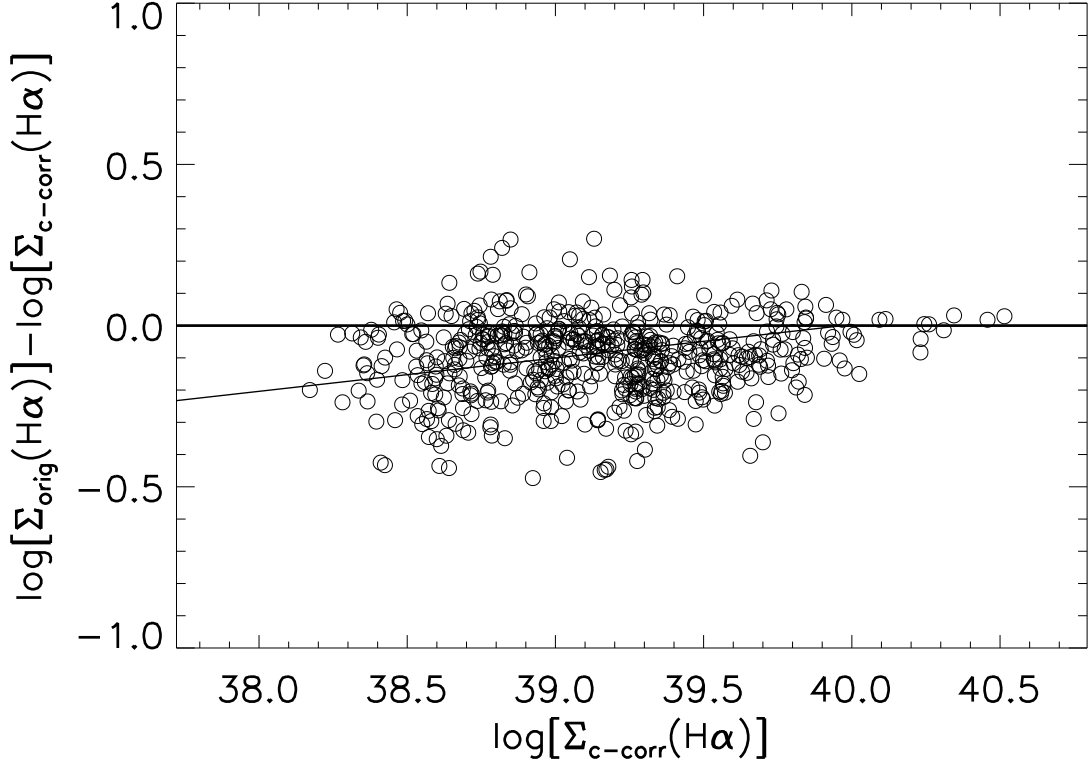}
}
\caption[]{ \centering The residual as a function of constant aperture correction corrected photometry is fitted with a linear line to establish the surface brightness dependent aperture correction. The convolved photometry data have been corrected for the constant aperture correction and the subscript, c-corr, is used to indicate the constant aperture correction corrected photometry here and in the paper. The units are $\rm ergs\cdot s^{-1}\cdot kpc^{-2}$ for LSDs.}
\label{fig:apc2}
\end{figure*}

After the constant aperture correction is applied, Figure \ref{fig:apc2} shows the difference between original and c-corr (constant-corrected, see section \ref{consapc}) photometry versus c-corr photometry. We change the x-axis from original photometry to the c-corr photometry because when we apply the surface brightness dependent aperture correction on the 24~$\micron$ and 70~$\micron$ data, we need to rely on the c-corr photometry as we have no access to the original high resolution photometry. We fit the data with $\rm{log}[\Sigma_{orig}(H\alpha)]<39.971$ with a linear function giving a best fit:

\begin{center}
$\rm{log}[\Sigma_{orig}(H\alpha)]-\rm{log}[\Sigma_{c-corr}(H\alpha)]=0.103[\rm{log}[\Sigma_{c-corr}(H\alpha)]-39.971]$
\end{center} 

as indicated by the black line in Figure \ref{fig:apc2}. Thus we have determined the empirical aperture correction for H$\alpha$ as two steps,

1) the constant aperture correction

\begin{equation}
\label{equ:ha_apc_1}
\Sigma_{c-corr}(H\alpha)=1.788\times \Sigma_{conv}(H\alpha)
\end{equation}
to correct the convolved photometry to the c-corr photometry and 

2) the surface brightness dependent aperture correction

\begin{equation}
\label{equ:ha_apc_2}
\Sigma_{l-corr}(H\alpha)=10^{0.103[\rm{log}[\Sigma_{c-corr}(H\alpha)]-39.971]}\times \Sigma_{c-corr}(H\alpha)
\end{equation}
for $\rm{log}[\Sigma_{c-corr}(H\alpha)]<39.971$ to correct the c-corr photometry to l-corr (luminosity-corrected) photometry.

\begin{figure*}
\center{
\includegraphics[scale=0.9]{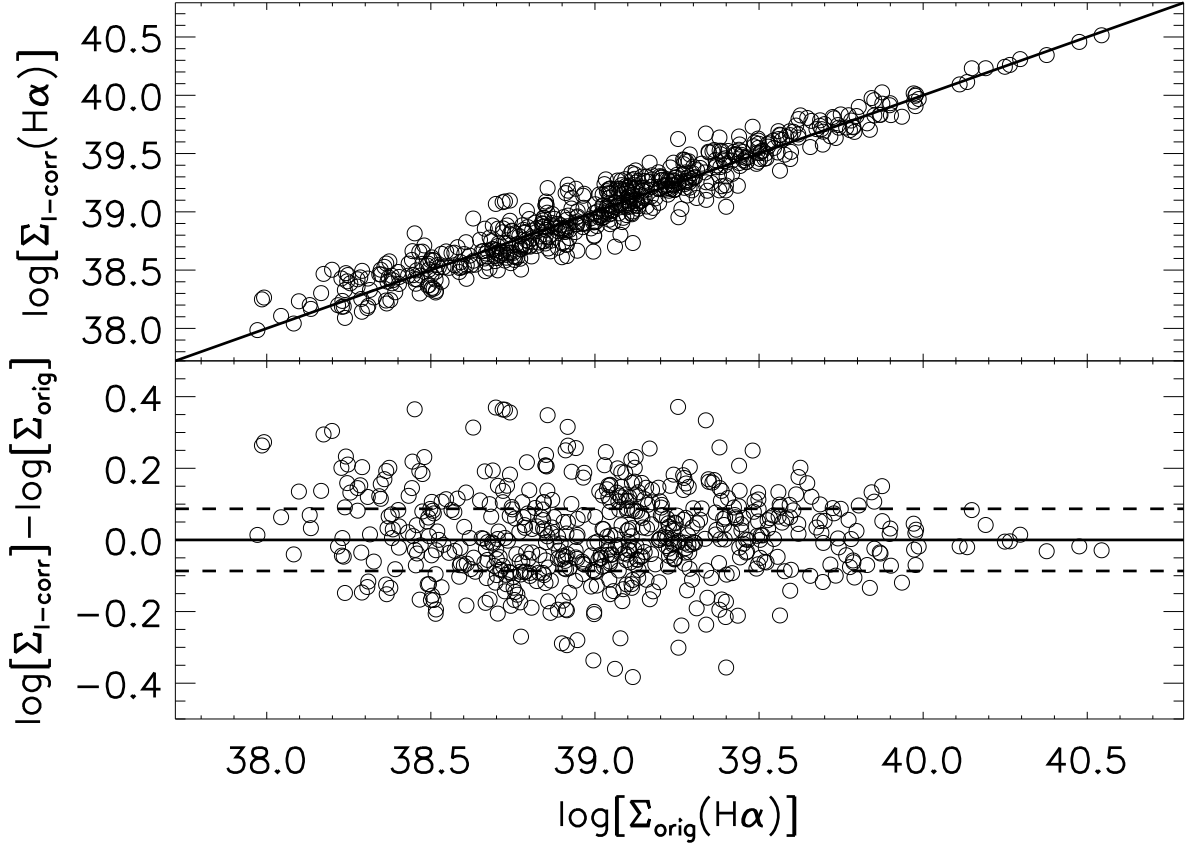}
}
\caption[]{ \centering The corrected photometry as a function of original H$\alpha$ photometry on the upper panel, and the residual as a function of original H$\alpha$ photometry on the lower panel, where the 1$\sigma$ dispersion (68\% data envelope, dashed lines) is $\sim0.09~dex$. The solid line indicates the one-to-one line. The subscript, l-corr, is used to indicate the surface brightness dependent aperture correction corrected photometry here and in the paper, which is also the two-step aperture correction corrected photometry and thus the proxy to the correct photometry. In the rest of the paper, all luminosity and LSD data is from the l-corr photometry unless otherwise specified. The units are $\rm ergs\cdot s^{-1}\cdot kpc^{-2}$ for LSDs.}
\label{fig:apc3}
\end{figure*}

Figure \ref{fig:apc3} shows the corrected (l-corr) photometry and the difference between the corrected (l-corr) and original photometry versus the original photometry. We estimate the $1-\sigma$ aperture correction uncertainty to be $\sim22.3\%$ by calculating the RMS of the deviation from the unity slope line (lower panel on Fig. \ref{fig:apc3}).

\begin{figure*}
\center{
\includegraphics[scale=0.9]{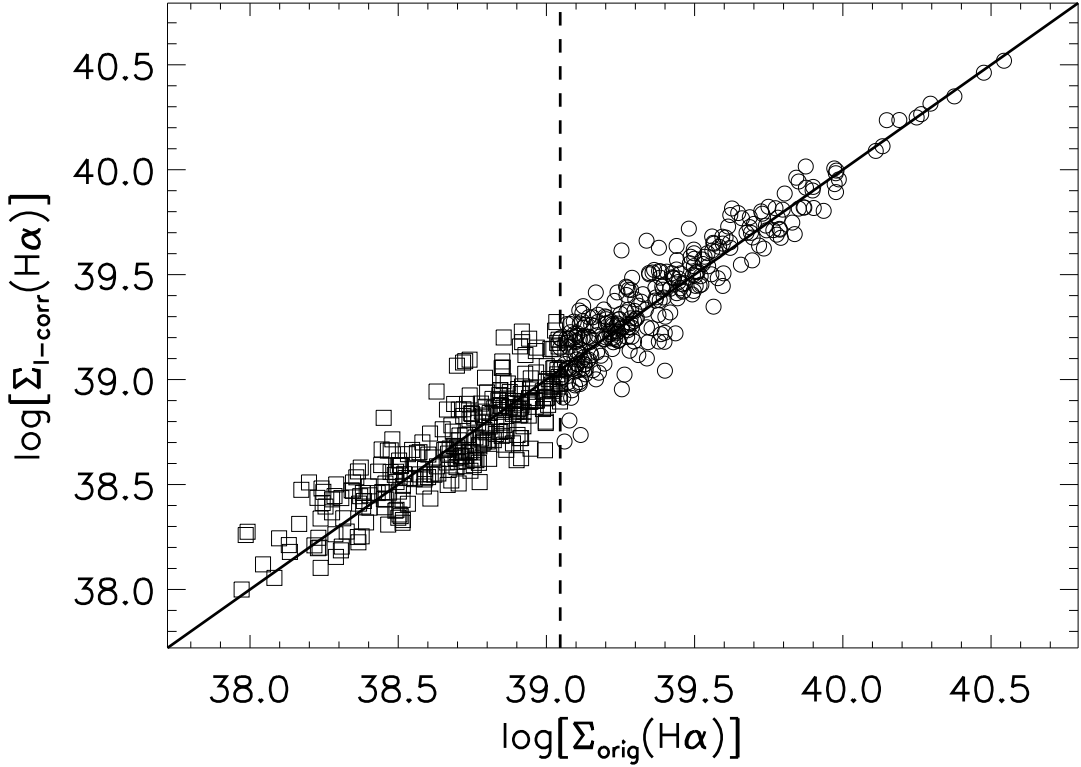}
}
\caption[]{ \centering The l-corr photometry as a function of the original H$\alpha$ photometry, while the aperture correction used here is established using only the brighter half, in LSD, of the data (circles right of the dashed line). The solid line indicates the one-to-one line and the dashed line separates the brighter half and the dimmer half. The squares show that the aperture correction established with high end half of data applies to the less luminous data as well, which suggests that our aperture corrections are robust. The units are $\rm ergs\cdot s^{-1}\cdot kpc^{-2}$ for LSDs.}
\label{fig:robust}
\end{figure*}

We further test the robustness of an aperture correction by considering only the brightest half of our sample (in terms of LSD) and repeating the above analysis. The results are shown in Figure \ref{fig:robust}, with the brightest half of the sample shown as circles. As can be seen from the Figure, the final results do not change whether the entire sample or the brightest half is used to establish the aperture correction. 

We then apply these aperture corrections to 70~$\micron$ and 24~$\micron$, according to a relation that $0.031\times L(24)\sim L(H\alpha)$ and $0.2\times L(70)\sim L(24)$ \citep{calzetti2007,calzetti2010}; an analysis of the distribution of 70~$\micron$ and 24~$\micron$ versus H$\alpha$ photometry on our data gives similar factors. The constant aperture correction remains the same for these two IR bands while the second step, Equation (\ref{equ:ha_apc_2}), becomes
\begin{equation}
\label{equ:24_apc}
\Sigma_{l-corr}(24)=10^{0.103[\rm{log}[0.031\Sigma_{c-corr}(24)]-39.971]}\times \Sigma_{c-corr}(24)
\end{equation}
for $\rm{log}[0.031\Sigma_{c-corr}(24)]<39.971$, and
\begin{equation}
\label{equ:70_apc}
\Sigma_{l-corr}(70)=10^{0.103[\rm{log}[0.2\times0.031\Sigma_{c-corr}(70)]-39.971]}\times \Sigma_{c-corr}(70)
\end{equation}
for $\rm{log}[0.031\times0.2\Sigma_{c-corr}(70)]<39.971$.

As we have no information about the original photometry of 70~$\micron$ and 24~$\micron$, the l-corr photometry will be a proxy to represent the true photometry of these two bands, and for consistency, we will also use the l-corr photometry of H$\alpha$ in the following analysis. All the luminosity and LSD values hereafter are derived from the l-corr photometry unless otherwise specified.

\section{Construction of the Simple Model}
\label{ap:model}

The simple model is constructed based on the model of \citet{calzetti2007}, and it has three basic ingredients: stellar population models \citep[STARBURST99,][]{leitherer1999}, dust attenuation \citep{calzetti2000}, and dust emission spectral energy distribution (SED) \citep[models from][]{draine2007a}.

For the stellar population models, we adopt a continuous star formation history based on the analysis of typical crossing time scale for our regions, calculated by using the sound speed and the aperture size, which is $\sim$ 100 Myr. We take this time scale as representative of our stellar population, but, for completeness, we generate constant star formation models between 2 Myr and 10 Gyr. For the stellar atmosphere metallicity provided by the STARBURST99 code, we choose Z=0.004, 0.02 and 0.05, which is roughly 0.3$Z_{\odot}$, 1.4$Z_{\odot}$ and 3.7$Z_{\odot}$, where $Z_{\odot}$ is equivalent to $12+\rm{log(O/H)}=8.69$ \citep{asplund2009}. We derive H$\alpha$ luminosities by integrating over the ionizing photon of the spectrum, and assuming case B recombination. After this step, we have a stellar population SED and the intrinsic H$\alpha$ luminosity.

Then we take the population SED and apply the attenuation curve by \citet{calzetti2000} to get the TIR emission, assuming all the absorbed UV-optical light has been reradiated in the IR. The E(B-V) value needed for the attenuation curve is calculated from the ionizing photon number using an empirical relation in \citet{calzetti2007}. The scaling of E(B-V) with metallicity is also taken into account \citep{calzetti2007}. For the nebular line attenuations, we adopt differential extinction with the stellar continuum (E$_{s}$=0.44E$_{g}$) as presented in \citet{calzetti2000} plus the Milky Way extinction curve. With the attenuation curve applied, we have the TIR luminosity and the observed H$\alpha$ luminosity.

We finally use the dust model from \citet{draine2007a} to determine the L24/TIR and L70/TIR fractions and to further get 24~$\micron$ luminosity and 70~$\micron$ luminosity. The dust emission models are parametrized as a function of $\rm q_{PAH}$, the fraction of PAH molecules, and U, the average stellar radiation field strength. $\rm q_{PAH}$ has only a modest impact on our output, and we adopt two values: $\rm q_{PAH}=4.6\%$ for solar and super-solar metallicity models, and $\rm q_{PAH}=0.47\%$ for the sub-solar metallicity model. We obtain estimates of U by integrating the stellar population SED using an approach similar to that presented in \citet{calzetti2007}. By combining these inputs with the models, we have the observed 24~$\micron$ luminosity and 70~$\micron$ luminosity.

Finally, we assume each model is observed within an unresolved 700pc aperture, which is the median physical size of all the apertures, and get the LSD values for H$\alpha$, 24~$\micron$ and 70~$\micron$ of the model, to compare with the data. The change in the adopted aperture for deriving model LSDs only causes a small shift (smaller than the dispersion of data) of the model lines. Also a test on the data of two fixed physical sizes ($\sim$ 250~pc and $\sim$ 650~pc with $\sim$ 100 apertures each) reveals that the correlations for data of difference physical sizes are consistent with each other.

\newpage

\begin{deluxetable}{cccrrccc} 
\tablecolumns{8}
\tablewidth{0pc}
\tablecaption{Sample \label{tab:sample}}
\tablehead{
\multicolumn{1}{c}{Galaxy} &
\multicolumn{1}{c}{Type} &
\multicolumn{1}{c}{Nuc} & 
\multicolumn{1}{c}{D} &
\multicolumn{1}{c}{Size} &
\multicolumn{1}{c}{Oxygen Abundance} &
\multicolumn{1}{c}{\# Regions} &
\multicolumn{1}{c}{Sub-sample} \\ 
 &
 &
 &
\multicolumn{1}{c}{(Mpc)} &
\multicolumn{1}{c}{(pc/16$''$)} &
\multicolumn{1}{c}{(Information)} &
& 
\\
\multicolumn{1}{c}{(1)} &
\multicolumn{1}{c}{(2)} &
\multicolumn{1}{c}{(3)} &
\multicolumn{1}{c}{(4)} &
\multicolumn{1}{c}{(5)} &
\multicolumn{1}{c}{(6)} &
\multicolumn{1}{c}{(7)} &
\multicolumn{1}{c}{(8)}
}
\startdata
NGC0024 & SAc & SF & 8.2 & 636 & 1 & 3 & M\\
NGC0337 & SBd & SF & 24.7 & 1916 & 1 & 5 & M \\
NGC0628 & SAc & SF & 11.4 & 884 & 2 & 40 & M \\
NGC0925 & SABd & AGN & 10.4 & 807 & 2 & 23 & M \\
NGC1097 & SBb & AGN & 16.9 & 1311 & 2 & 20 & M \\
NGC1512 & SBab & AGN & 10.4 & 807 & 1 & 7 & H \\
NGC1566 & SABbc & AGN & 18 & 1396 & 3 & 15 & H \\
NGC1705 & Am & SF & 5.8 & 450 & 1 & 2 & L \\
NGC2403 & SABcd & SF & 3.5 & 271 & 2 & 41 & M \\
Ho II & Im & SF & 3.5 & 271 & 1 & 8 & L \\
NGC2798 & SBa & AGN & 24.7 & 1916 & 1 & 1 & M \\
NGC2841 & SAb & AGN & 9.8 & 760 & 2 & 9 & H \\
NGC2976 & SAc & SF & 3.5 & 271 & 1 & 9 & M \\
NGC3049 & SBab & SF & 19.6 & 1526 & 1 & 1 & H \\
NGC3190 & SAap & AGN & 17.4 & 1350 & 3 & 3 & H \\
NGC3184 & SABcd & SF & 8.6 & 667 & 2 & 27 & H \\
IC2574 & SABm & SF & 3.5 & 271 & 1 & 10 & L \\
Mrk33 & Im & SF & 21.7 & 1683 & 1 & 1 & M \\
NGC3351 & SBb & SF & 9.3 & 721 & 2 & 8 & H \\
NGC3521 & SABbc & AGN & 9 & 698 & 2 & 22 & M \\
NGC3627 & SABb & AGN & 8.9 & 690 & 1 & 8 & M \\
NGC3938 & SAc & SF & 12.2 & 946 & 3 & 22 & M \\
NGC4254 & SAc & SF & 20 & 1551 & 2 & 20 & M \\
NGC4321 & SABbc & AGN & 20 & 1551 & 2 & 24 & H \\
NGC4450 & SAab & AGN & 20 & 1551 & 3 & 5 & H \\
NGC4536 & SABbc & SF/AGN & 25 & 1939 & 1 & 8 & M \\
NGC4559 & SABcd & SF & 11.6 & 900 & 2 & 17 & M \\
NGC4579 & SABb & AGN & 20 & 1551 & 3 & 7 & H \\
NGC4631 & SBd & SF & 9 & 698 & 1 & 14 & M \\
NGC4725 & SABab & AGN & 17.1 & 1326 & 1 & 22 & M \\
NGC5055 & SAbc & AGN & 8.2 & 636 & 2 & 24 & M \\
NGC5194 & SABbc & AGN & 8.2 & 636 & 2 & 35 & H \\
NGC5474 & SAcd & SF/AGN & 6.9 & 535 & 1 & 6 & M \\
NGC5713 & SABbcp & SF & 26.6 & 2063 & 1 & 1 & M \\
IC4710 & SBm & SF & 8.5 & 659 & 3 & 8 & M \\
NGC6822 & IBm & SF & 0.6 & 47 & 1 & 18 & L \\
NGC6946 & SABcd & SF & 5.5 & 427 & 2 & 37 & M \\
NGC7331 & SAb & AGN & 15.7 & 1218 & 2 & 22 & M \\
NGC7552 & SAc & SF & 22.3 & 1730 & 1 & 3 & M \\
NGC7793 & SAd & SF & 3.2 & 248 & 2 & 41 & M \\
\enddata 
\tablerefs{(1) \citet{kennicutt2003}; (2) \citet{moustakas2010}.}
\tablecomments{Col. (1) SINGS galaxy name; Col. (2) Morphological type; Col. (3) Adopted nuclear optical spectral classification from \citet{moustakas2010}; type AGN is adopted when it is SF/AGN in \citet{moustakas2010}; Col. (4) Distance; morphological type and distance are adopted from \citet{kennicutt2003} and also listed in \citet{moustakas2010}; Col. (5) Physical size of adopted aperture; Col. (6) Oxygen abundance information adopted from \citet{moustakas2010}; 1 = characteristic value for the galaxy is adopted; 2 = gradient is adopted and metallicity calculated for each aperture; 3 = L-Z relation derived value is adopted; Col. (7) The number of apertures selected from each galaxy; Col. (8) The sub-sample each galaxy belongs to, but the regions in the galaxy may belong to another sub-sample according to metallicity gradients, if available; L = low metallicity sample; M = intermediate metallicity sample; H = high metallicity sample.}
\end{deluxetable}

\newpage

\clearpage

\end{document}